\documentclass[aps,prd,preprint,nofootinbib,showpacs]{revtex4-1}
\usepackage{amsmath,amssymb}
\usepackage{url,color}
\usepackage{graphicx}
\usepackage{slashed}
\usepackage{dsfont}
\usepackage{float}
\usepackage{multirow} 
\usepackage{booktabs} 
\usepackage{pifont} 
\usepackage{subfigure}
\usepackage{caption}
\makeatletter
\newif\if@preliminary
\@preliminaryfalse
\def\preliminary{\@preliminarytrue}
\def\preprintno#1{\def\@preprintno{#1}}
\def\address#1{\def\@address{#1}}
\def\email#1#2{\thanks{\tt #1@{}#2}}
\long\def\@makecaption#1#2{%
  \vskip\abovecaptionskip
  \sbox\@tempboxa{#1: \emph{#2}}%
  \ifdim \wd\@tempboxa >\hsize
    #1: \emph{#2}\par
  \else
    \hbox to\hsize{\hfil\box\@tempboxa\hfil}%
  \fi
  \vskip\belowcaptionskip}
\makeatother

\let\Re\relax
\DeclareMathOperator{\Re}{Re}
\let\Im\relax
\DeclareMathOperator{\Im}{Im}
\newcommand{\ii}{\mathrm{i}}
\newcommand{\ee}{\mathrm{e}}
\newcommand{\dd}{\mathrm{d}}

\newcommand{\LL}{\mathcal{L}}

\newcommand{\vV}{\mathbf{V}}

\newcommand{\vW}{\mathbf{W}}

\newcommand{\vH}{\mathbf{H}}
\newcommand{\Tr}{\mathop{\rm Tr}}
\newcommand{\tr}[1]{\operatorname{tr}\left[#1\right]}

\newcommand{\vB}{\mathbf{B}}
\newcommand{\vD}{\mathbf{D}}

\newcommand{\GeV}{\text{GeV}}
\newcommand{\TeV}{\text{TeV}}

\newcommand{\amp}{\mathcal{A}}
\newcommand{\bss}{\begin{tiny}}
\newcommand{\ess}{\end{tiny}}
\begin{document}


\preprint{SI-HEP-2014-20}
\preprint{DESY-14-142}

\title{%
  High-Energy Vector Boson Scattering \\
  after the Higgs Discovery
}

\author{Wolfgang Kilian}
\email{kilian}{physik.uni-siegen.de}
\affiliation{Department of Physics, %
  University of Siegen, %
  D--57068 Siegen, Germany} 
\author{Thorsten Ohl}
\email{ohl}{physik.uni-wuerzburg.de}
\affiliation{Faculty of Physics and Astronomy, %
    W\"urzburg University, %
    D--97074 W\"urzburg, Germany}
\author{J\"urgen Reuter}
\email{juergen.reuter}{desy.de}
\affiliation{DESY Theory Group, %
    D--22603 Hamburg, Germany}
\author{Marco Sekulla}
\email{sekulla}{physik.uni-siegen.de}
\affiliation{Department of Physics, %
  University of Siegen, %
  D--57068 Siegen, Germany}

\date{\today}

\begin{abstract}
  Weak vector-boson $W,Z$ scattering at high energy
  probes the Higgs sector and is most sensitive to any new
  physics associated with electroweak symmetry breaking.  We show that
  in the presence of the $125\;\GeV$ Higgs boson,
  a conventional effective-theory analysis fails for this class of
  processes.  We propose to extrapolate the effective-theory ansatz by an
  extension of the parameter-free \textit{K}-matrix unitarization
  prescription, which we denote as direct \textit{T}-matrix
  unitarization.  We generalize this prescription to arbitrary
  non-perturbative models and describe the 
  implementation, as an asymptotically consistent reference model
  matched to the low-energy effective theory.  We present exemplary
  numerical results for full six-fermion processes at the LHC.
\end{abstract}


\maketitle


\section{Introduction}
\label{sec:intro}

After the discovery of a Higgs-like particle at the
LHC~\cite{Aad:2012tfa,Chatrchyan:2012ufa}, and without any
signal of other new particles, the focus of collider physics is shifting
towards a detailed study of electroweak symmetry breaking
(EWSB). We are interested in the properties of the Higgs boson itself
and in its precise role in a fundamental
theory~\cite{ESG2013,Brock:2014tja,Degrande:2013rea}. Beyond that,
the most fundamental process of the electroweak interactions is the scattering
of 
the electroweak gauge bosons~\cite{Gianotti:2014}. It will be one of
the key physics processes at the high-luminosity LHC as well as any
planned future high-energy $pp$ and $e^+e^-$ machine.

The most striking effect of the Higgs boson is the strong suppression
of electroweak vector-boson scattering (VBS) at high center-of-mass (c.m.)
energy~\cite{Lee:1977yc}.  Without the Higgs boson, VBS scattering
amplitudes $VV\to VV$, where $V=W^\pm$ or $Z$, are dominated by scalar
Goldstone-boson scattering which relates to the scattering of
longitudinally polarized $W$ and $Z$ particles.  Power counting
predicts an $s/v^2$ rise of these amplitudes ($v=(\sqrt2\,
G_F)^{-1/2}=246\,\GeV$), such that electroweak interactions should
become strong in the TeV range.  However, the Standard Model (SM) representation of
the Higgs sector replaces this by a consistently weakly interacting model.  The
cancellation induced by Higgs exchange results in a residual
Goldstone-scattering amplitude that is asymptotically small, at tree
level proportional to $m_H^2/v^2 = 0.25$.  This can be interpreted as
an effective suppression in the cross section which for a $VV$ c.m.\
energy of $\sqrt{s}=1.2\;\TeV$ amounts to a factor of
$m_H^4/s^2=10^{-4}$.

At the LHC, VBS processes have become accessible to
experiment~\cite{Aad:2014zda,CMS:2014uib}.  
The accuracy and energy reach of these
measurements will improve at the upgraded LHC and at future colliders,
including the planned ILC~\cite{Beyer:2006hx}.  The SM with the
observed light Higgs particle provides a very specific prediction for
all VBS processes, namely a scattering amplitude which is dominated by
the transversal gauge-boson components of the $W$ and $Z$ bosons.  A
significant excess in the longitudinally polarized channel would
clearly point to new interactions in the EWSB sector. 

A phenomenological description of high-energy VBS processes should
smoothly interpolate between the low-energy behavior, which is
determined by the SM and depends on a well-defined set of perturbative
parameters as corrections, and any possible high-energy asymptotics
which should be captured by a sufficiently generic class of models~\cite{Reuter:2013gla,Reuter:2014kya}.
It is important to note that in hadron collider observables, the
separation of low- and high-energy scattering is not straightforward.
For a meaningful comparison with data, the parameterized high-energy
behavior has to remain consistent with the universal principles of
quantum physics.  Systematically comparing model predictions with
data, the results will become a measure of confidence for the SM case,
or otherwise the numerical evaluation of any observed new-physics
effects.

In this paper, we develop this program specifically for the scenario with a
light Higgs boson which is now being confirmed by the LHC analyses.  This
scenario deviates significantly from the situation without light
Higgs~\cite{Chanowitz:1984ne,Chanowitz:1985hj,Chanowitz:1986hu,Chanowitz:1987vj,Chanowitz:1993zh,Alboteanu:2008my}
where there is a steady transition from low-energy 
weak interactions to strong interactions at high energies.  We discuss the
necessary steps that allow us to parameterize high-energy asymptotics and the
interpolation between low and high energies, embed this in the interacting
theory with off-shell gauge bosons and fermions, and show how to convert the
algorithm into a consistent calculational method and simulation of exclusive
event samples.

The paper consists of three parts.  In the first part, we review the
essentials of the effective-theory approach to electroweak
interactions and the Higgs mechanism.  The second part extends the
well-known concept of \textit{K}-matrix unitarization in such a way that we
can apply it to generic (non-Hermitian) expansions and models of the
complete scattering matrix.  In the third part, we show how to implement
this variant of \textit{K}-matrix unitarization in actual calculations of vector-boson
scattering amplitudes beyond the Standard Model and show exemplary
numerical results for LHC processes.  In a final section, we summarize
the results and conclude.

\section{Effective Theories for Electroweak Interactions}

\subsection{Effective Theory and Higgs Mechanism}

Throughout this paper, we will assume that no new weakly coupled new
particles, i.e., narrow resonances, appear within the energy range
that we consider for VBS.  The elementary particle spectrum is given
by the SM.  It has been known for a long time that this scenario can
be addressed by an effective field theory (EFT) as a universal phenomenological ansatz~\cite{Weinberg:1968de}.  

Early studies of VBS considered a nonlinear EWSB representation, the
chiral electroweak Lagrangian, as an EFT without light Higgs boson~\cite{Appelquist:1980vg,Longhitano:1980iz,Dawson:1990cc,Appelquist:1993ka,Dobado:1995qy,Dobado:1995ze,Dobado:1997jx,Kilian:2003pc,Boos:1997gw,Boos:1999kj,Belyaev:1998ih,Buchalla:2012qq}.
This scenario was to be experimentally distinguished from the simplest light-Higgs
case~\cite{Gupta:1993tq,Chanowitz:1993zh,Bagger:1993zf,Bagger:1995mk,Barger:1995cn,Gupta:1995ru}.  Any Higgs-less model
evolves into strong interactions in the TeV range, while the SM
remains weakly interacting at all energies.  However, after the recent
discovery of a light Higgs
candidate~\cite{Aad:2012tfa,Chatrchyan:2012ufa}, new studies should
narrow down the 
case towards distinguishing different models which do include the
Higgs as a particle.

A neutral scalar particle can be coupled to the nonlinear chiral
Lagrangian in a gauge-invariant way, including a power series of
higher-dimensional operators~\cite{Feruglio:1992wf,Grinstein:2007iv,Alonso:2012px,Alonso:2012pz,Buchalla:2013rka}. Alternatively, we can combine it with
the Goldstone bosons of EWSB as an electroweak doublet and base the
analysis on the SM, also augmented by a power series of
higher-dimensional operators~\cite{Buchmuller:1985jz,Hagiwara:1992eh,Hagiwara:1993ck,Grzadkowski:2010es}.  Both approaches allow for the most
general set of interactions.  They are related by a simple field
redefinition and thus equivalent~\cite{Cornwall:1973tb,Cornwall:1974km,Bergere:1975tr,Weinberg:1978kz,Kilian:1998bh,Kilian:2003pc}.  However, truncating either power
series exposes differences in the power counting, and thus
different theoretical prejudice about the hierarchy of coefficients.

In this work, we anticipate Higgs (and $W,Z$) couplings that are close
to their SM values, as suggested by the current LHC
analyses~\cite{ICHEP2014}.  In the
linear representation, this parameter point is distinguished by
renormalizablity, the absence of any higher-dimensional terms.  In the
nonlinear representation this parameter point is not distinguished in
the Lagrangian, so the high-energy cancellations that the Higgs
induces at the amplitude level
appear as accidental.  We therefore adopt the linear representation.
Furthermore, we implicitly assume that electroweak gauge symmetry is a
meaningful concept up to energies far beyond the TeV scale~\cite{Cornwall:1973tb,Cornwall:1974km}.  We
therefore include the gauge boson fields $W_\mu^{1,2,3}$ and $B_\mu$
as elementary vector fields which enter via covariant derivatives and
field strength tensors, always multiplied by the respective gauge
couplings $g$ and $g'$ and thus weakly interacting.  This assumption
is clearly supported by all known electroweak precision and flavor
data.

The EFT extension of the linearily parameterized SM has been worked
out up to next-to-leading order in the power series
(dimension six)~\cite{Buchmuller:1985jz,Hagiwara:1993ck,Grzadkowski:2010es,Passarino:2012cb}
and applied to properties of the Higgs boson in various contexts~\cite{Kilian:1996wu,Kilian:2003xt,Giudice:2007fh,Espinosa:2010vn,Boos:2013mqa,Contino:2013kra,Hagiwara:1993qt,Alam:1997nk,Grojean:2013kd,Belusca-Maito:2014dpa,Biekoetter:2014jwa}.  Operator mixing at
the one-loop order has been calculated in
Refs.~\cite{Elias-Miro:2013gya,Elias-Miro:2013mua,Jenkins:2013zja,Jenkins:2013wua,Alonso:2013hga}.
Dimension-eight operators as the second order have been 
studied in
Refs.~\cite{Eboli:2006wa,Degrande:2013kka}.  In the current work, we
do not intend to incorporate the complete operator basis, but rather
select exemplary terms that specifically affect VBS, such that we can
describe the matching and interpolation procedure that connects low-
and high-energy amplitudes.

\subsection{Fields and Operators}

The SM Higgs resides in a doublet of the $SU(2)_L$ gauge symmetry.
Our notation is laid out in Appendix~\ref{sec:notation}.  We choose to
parameterize the Higgs multiplet in form of a $2\times 2$ Hermitian
matrix $\vH$.  In this parameterization, the custodial-$SU(2)_C$
transformation properties of any operator are manifest, and there is a
simple relation to the nonlinear Higgs EFT, namely the replacement
\begin{equation}
  \vH \to \frac{1}{2}(v + h)\Sigma
\end{equation}
where $\Sigma$ is a nonlinear Goldstone-boson representation.

Since we focus exclusively on the Higgs and electroweak gauge
sectors, we do not write light fermions explicitly, but treat them as
external probes for the interactions that we are interested in.  In
accordance with the hypothesis of minimal flavor violation, we ignore
the possibility of anomalous effects due to higher-dimensional
operators that involve light flavors.  Heavy flavors and gluons do not
play a role for the signal processes that we consider. If we do not
look at observables with explicit heavy flavors, the fermion sector
emerges as perturbative.  Extending this result to the full EFT,
we arrive at a model that decomposes, at high energy $E\gg v$, into
left- and right-handed fermion, gauge boson, and scalar (EWSB)
sectors, almost mutually decoupled due to the smallness of the EWSB
order parameter $v$.  This decomposition is stable against radiative
corrections, since operator mixing in the EFT is governed exclusively
by weak couplings with loop factors.  It should be noted that it is
also stable with respect to applying equations of motion to the
operator basis, as long as we impose the gauge and minimal flavor
violation principles that identify weak coupling parts.

The processes of interest at a hadron collider, namely
\begin{align}
\label{VBS-full}
  pp &\to 2j + (VV \to 4f)
\end{align}
embed the actual quasi-elastic VBS processes, $VV\to VV$, together
with irreducible non-VBS background.  The vector-boson interactions
are affected by all bosonic dimension-six and dimension-eight
operators that the EFT provides.  We should weigh their impact in view
of the experimental possibilities.  Current and future analyses will
rather precisely determine the coefficients of pure-gauge operators
that affect vector-boson pair production and related processes.
Fixing a suitable operator basis, we may take these coefficients as
given~\cite{Corbett:2013pja,Pomarol:2013zra}.  On the other hand, we can safely ignore terms that exclusively
provide couplings to Higgs pairs, since such couplings do not enter
VBS processes at tree level.  In a simplified first approach to the
problem, we may thus exclude most dimension-six
operators from an analysis that focuses on VBS.  Instead, we
incorporate operators that supply genuine quartic gauge couplings in
the longitudinal mode.  Such operators do not affect simpler
processes, they occur first at dimension eight in the operator basis.

For the purpose of studying VBS processes, we therefore concentrate on
the subset
\begin{alignat}{3}
  \label{LL-HD}
  \LL_{HD} &=
  & & F_{HD}\ &&
  \tr{{\vH^\dagger\vH}- \frac{v^2}{4}}\cdot
  \tr{\left (\vD_\mu \vH \right )^\dagger \left (\vD^\mu \vH \right )}
\\
  \label{LL-S0}
	\LL_{S,0}&=
	 & &F_{S,0}\ &&
	  \tr{ \left ( \vD_\mu \vH \right )^\dagger \vD_\nu \vH}
		\cdot \tr{ \left ( \vD^\mu \vH \right )^\dagger \vD^\nu \vH}
\\
  \label{LL-S1}
	\LL_{S,1}&=
	 & &F_{S,1}\ &&
	  \tr{ \left ( \vD_\mu \vH \right )^\dagger \vD^\mu \vH}
		\cdot \tr{ \left ( \vD_\nu \vH \right )^\dagger \vD^\nu \vH} 
\end{alignat}
The corresponding Feynman rules modify the VBS amplitude expressions,
predominantly in the longitudinally polarized channel.

The dimension-six operator $\LL_{HD}$ modifies the $HWW$ and $HZZ$
couplings and thus controls the Higgs exchange diagrams in VBS. We
take this particular term as a representative of the possible effects
that dimension-six operators can contribute to VBS processes. We have
written the operator in a subtracted form, such that it respects
on-shell renormalization conditions as discussed in
App.~\ref{sec:notation}.

The included terms manifestly respect custodial symmetry,
$SU(2)_C$~\cite{Sikivie:1980hm}.  There are both dimension-six and
dimension-eight operators that violate $SU(2)_C$, but they provide
bilinear and trilinear gauge couplings and thus should be considered
as input to a VBS analysis.  $SU(2)_C$-violating operators which only
affect quartic couplings occur first at dimension 10.  This is a
consequence of the linear doublet Higgs representation.  We therefore assume
global $SU(2)_C$ invariance for the current paper, which should hold
at least at the threshold where new effects start to become relevant.

\subsection{Breakdown of the EFT}

The pure-SM cross section for VBS, (\ref{VBS-full}), is dominated by
transversally polarized gauge bosons, which in the high-energy limit
decouple from the Higgs sector.  Apart from the Higgs suppression,
this is a consequence of the vector-boson production mechanism, namely
radiation from massless fermions which couple to longitudinal vector
bosons only via helicity mixing~\cite{Kane:1984bb,Dawson:1984gx}.  The
transversal polarization directions are further enhanced by their
higher multiplicity.

\begin{figure}[htb]
	\centering
	\includegraphics[width=0.7\linewidth]{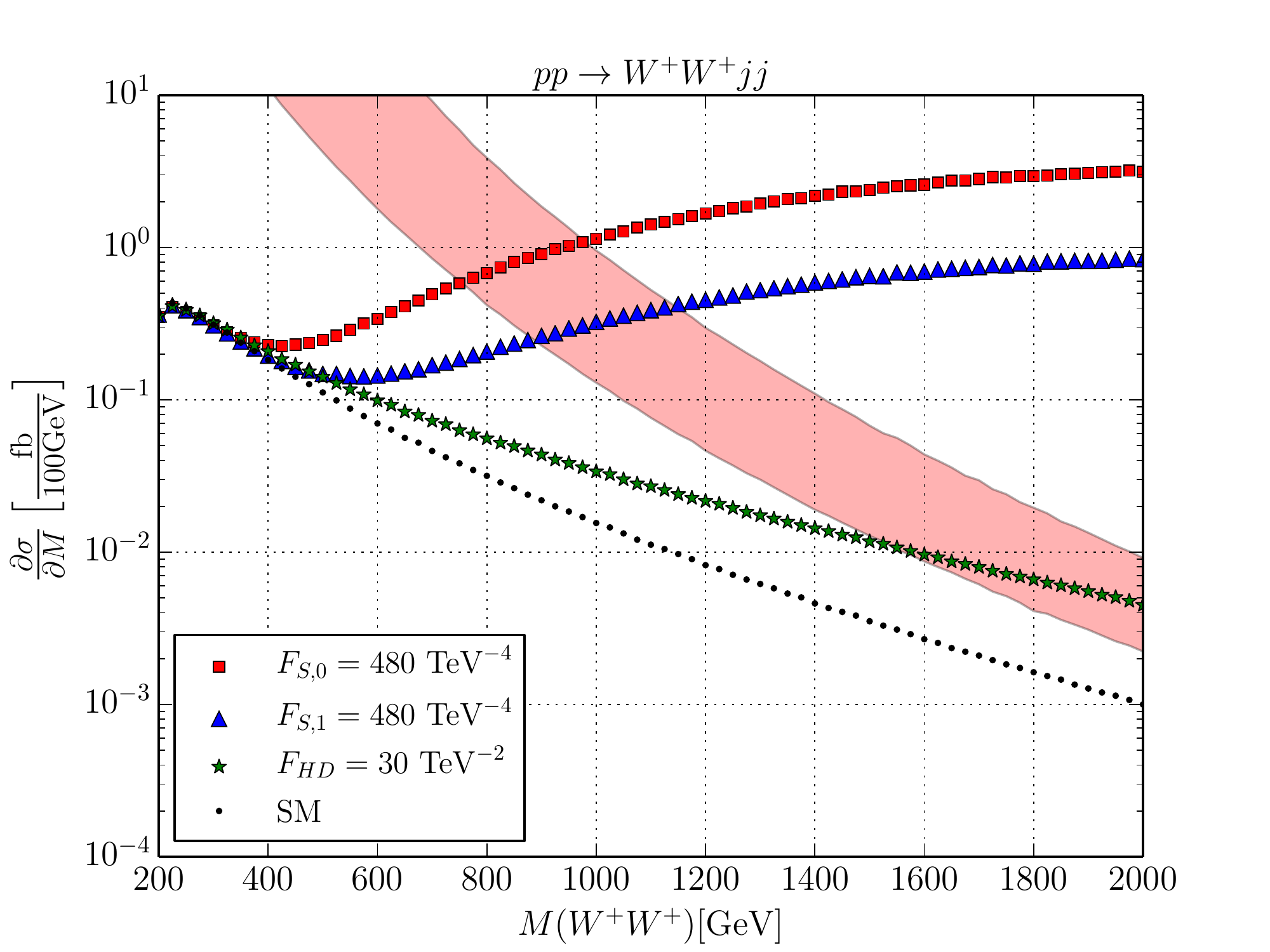}
  \caption{ $pp\to W^+W^+jj$, naive EFT results that violate
    unitarity, QCD contributions neglected. The band describes maximal
    allowed values, due to unitarity constraints, for the
    differential cross section. The lower bound describes the
    saturation of $\mathcal{A}_{20}$ and the upper bound describes the
    simultaneous saturation of $\mathcal{A}_{20}$ and
    $\mathcal{A}_{22}$, cf.~\eqref{eq:isospin-spin}. \\ 
    Cuts: $M_{jj} > 500$ GeV; 
    $\Delta\eta_{jj} > 2.4$;
    $p^j_T > 20$ GeV;
    $|\eta_j| > 4.5$.}
   \label{fig:W+W+bare}
\end{figure}

Adding in the operators (\ref{LL-HD})--(\ref{LL-S1}), the picture changes.
In Fig.~\ref{fig:W+W+bare}, we illustrate this for the
particular process of same-sign $W$ production at a LHC energy of
$14\,\TeV$.  We have applied standard cuts~\cite{Aad:2014zda} on the
forward jets and the $VV$ system, adapted to the simplified picture of
on-shell vector bosons in the final state.

For this figure, we have computed the complete process $pp\to
  W^+W^+ jj$ at leading order.  We used the Monte-Carlo integrator and
  event generator
  WHIZARD~\cite{Kilian:2007gr,Moretti:2001zz,Kilian:2011ka} with the
  CTEQ6L PDF set.  The SM curve is compared to three
  curves for models which contain a single nonzero coefficient for the
  effective higher-dimensional operators~(\ref{LL-HD}, \ref{LL-S0},
  \ref{LL-S1}), respectively, without any unitarization correction.
  For an indication of the unitarity limits, we have included a
  quartic Goldstone interaction amplitude with a constant coefficient
  $a_{IJ}=\ii$ in the $I=2$ and $J=0,2$ channels and recomputed the
  process with this modification.  The variation in the unitarity
  bound corresponds to the choice of saturating only one or both of
  these contributions.  This amplitude has been extended to physical
  vector bosons at finite energy and evaluated for off-shell
  initial-state vector bosons, according to the prescription that we
  describe below in Sec.~\ref{sec:completeEW}.  Due to the inherent
  ambiguities in such a prescription for finite energy, it is not
  possible to precisely state the unitarity limits for a physical
  cross section.  Nevertheless, we should constrain the validity
  region of the effective theory, given the chosen parameter values,
  to the energy range where the unitarity band is not yet touched by the
  corresponding curve.

The cross section with a dimension-six correction included, asymptotically falls
off with a slower rate than the SM reference curve.  There is a range
of coefficient values for which the EFT remains valid, until it
eventually crosses the unitarity bound. Looking at
Fig.~\ref{fig:W+W+bare}, we
observe that for the chosen coefficient value, unitarity can be
regarded as (marginally) satisfied, if we account for the limited
event count in an actual analysis which makes the lower part of the
diagram inaccessible.  For larger coefficient values, we would leave
the applicability range of the EFT.  This result
is typical for the effect of dimension-six operators in
energy-dependent observables~\cite{Degrande:2012wf,Biekoetter:2014jwa}.

By contrast, the dimension-eight operators have a dramatic impact on
the VV pair invariant-mass distribution.  The differential cross
section leaves the SM value at a certain threshold energy and then
\emph{increases} up to a broad maximum at multi-TeV invariant mass.
This behavior is easily explained by the high mass dimension of the
included operators.  Their contributions are enhanced by
$M_{WW}^8/m_H^8$ relative to the SM prediction.  The high power of
$M_{WW}$ overcomes the energy-dependent suppression caused by the
parton distributions. Taken at face value, this would become a
powerful handle on the coefficients $F_{S,0}$ and $F_{S,1}$, even for
a rather low collider luminosity.

Unfortunately, this result is entirely unphysical.  No high-energy
completion of the SM that is consistent with the basic assumptions of
the EFT approach is capable of producing such a distribution~\cite{Lee:1977yc}.  In the
dimension-eight case, the calculated curves cross this unitarity limit
immediately within the experimentally accessible region, for
\emph{any} coefficient value that could possibly be accessible.
Furthermore, except for the rare final state $ZZ\to 4\ell$,
observables at a hadron collider mix
different $M_{WW}$ ranges and thus disallow a strict exclusion of the
unphysical region in an analysis.

Obviously, we are using the EFT far beyond its region of validity.
The important result is that for the dimension-eight operators, which
are the most interesting terms in this context, there is actually
\emph{no} coefficient value for which the EFT yields a useful
prediction.  This is in contrast to an analysis of dimension-six
operators, which are mostly accessible via production and decay
processes with well-defined or limited energy range.  In other words,
if a deviation from the SM in VBS can be detected at all, it either
contains new particles which invalidate the SM-based EFT, or it
contains strong interactions.  In either case, the pure EFT is
insufficient.

\section{Unitarization Prescriptions}

\subsection{\textit{K}-Matrix Ansatz, Cayley Transform and Stereographic Projection}

To address the invalid high-energy asymptotics of an EFT in a
universal way, we start with the \textit{K}-matrix ansatz.  The formalism
applies to the complete $S$ matrix, so it is independent of any
particular model or approximation, and it does not rely on a
perturbative expansion.  It is therefore a suitable ansatz for the
present problem where we have no clue about the fundamental theory
that describes electroweak interactions, unless it is just the
Standard Model or a simple weakly interacting extension.

Heitler~\cite{Heitler:1941} and Schwinger~\cite{Schwinger:1948yk}
introduced the \textit{K} operator as the Cayley transform of the complete
unitary scattering operator $S$, namely
\begin{subequations}
\label{eq:Cayley}
\begin{equation}
  S = \frac{\mathbf{1}+\ii K/2}{\mathbf{1} - \ii K/2}\,,
\end{equation}
where we include a factor~$1/2$ for later convenience.  \textit{K} is
self-adjoint by definition, and as such more closely related to the
interaction Hamiltonian than the $S$~matrix.  The corresponding
transition operator~\textit{T}, as defined by~$S=\mathbf{1}+\ii T$, is then
\begin{equation}
\label{eq:T(K)}
  T = \frac{K}{\mathbf{1} - \ii K/2}\,.
\end{equation}
\end{subequations}
This \textit{T} satisfies the optical theorem $\ii T^\dagger T =
T-T^\dagger$ since $S$ is unitary, $SS^\dagger=S^\dagger S=\mathbf{1}$.

These relations can be inverted
\begin{equation}
\label{eq:K(T)}
  K = 2\ii \frac{1-S}{1+S} = \frac{T}{1+\ii T/2}\,.
\end{equation}
If the theory admits a perturbative expansion, the latter formula
allows us to compute the \textit{K}-matrix perturbatively from the expansion
of~\textit{T}, as long as~$T-2\ii$ is non-singular.  Obviously, $K=T$ in
lowest order.

\begin{figure}[hbt]
\begin{center}
  \includegraphics[width=0.8 \textwidth]{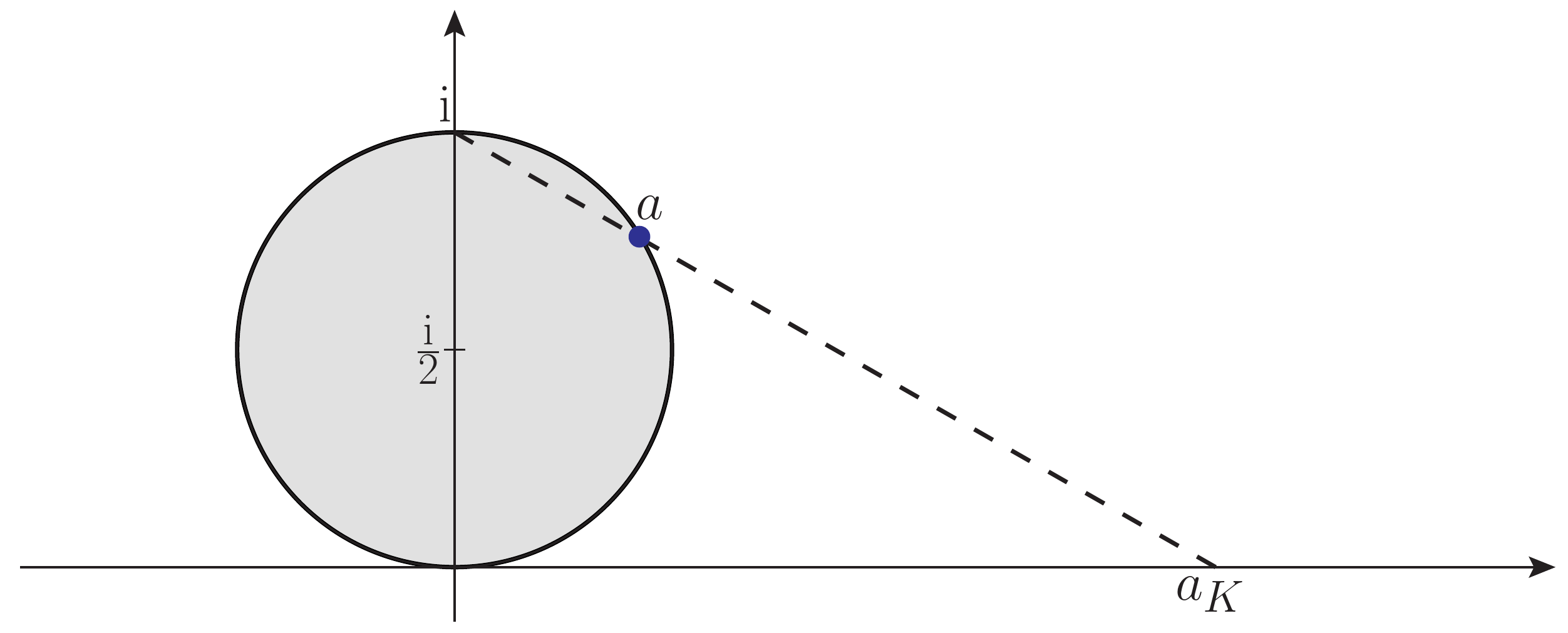}
\end{center}
\caption{Stereographic projection of a real scattering
  amplitude (\textit{K}-matrix eigenvalue) onto the Argand circle}
\label{fig:Argand-K-matrix}
\end{figure}

If we are able to find a basis that diagonalizes the scattering
operator $S$, and thus \textit{T} and \textit{K}, the Cayley transform has a simple
geometric interpretation for the eigenvalues.  Given a complex
eigenvalue $t=2a$ of the true transition operator \textit{T}, the optical
theorem implies
\begin{equation}
\label{eq:Argand}
  |a-\ii/2| = 1/2\,,
\end{equation}
i.e., the eigenamplitude $a$ is located on the Argand circle
with radius $1/2$ and center $\ii/2$~\cite{Beringer:1900zz}.  The
corresponding real \textit{K}-matrix eigenvalue $k=2a_K$ is then given by
\begin{equation}
\label{eq:aK(a)}
  a_K = \frac{a}{1 + ia}
\end{equation}
This is the inverse of the stereographic projection from the real axis
onto the Argand circle, cf. Fig.~\ref{fig:Argand-K-matrix}.  The
Cayley transform, or \textit{K}-matrix, can thus be understood as the
inverse stereographic projection of the transition
matrix \textit{T} onto the space of Hermitian matrices.

The scattering amplitude of charged particles will contain a Coulomb
singularity.  This singularity is physical and must not be handled by
an ad-hoc unitarization prescription, but by a proper definition of
the asymptotic states of charged
particles~\cite{Bloch:1937pw,Yennie:1961ad,Kulish:1970ut,Dahmen:1990cd}
instead.  Thus one should subtract the Coulomb singularity from the
amplitude, apply the chosen unitarization prescription to the
remainder and subsequently add the Coulomb singularity together with
appropriate corrections for the asymptotic states.

In the following, we will use the terms scattering operators and
scattering matrices interchangably.  We stress that we are always
dealing with the full $2\to n$-particle scattering operators.
Nevertheless, we may assume that we work in the finite dimensional
subspaces corresponding to a fixed overall angular momentum in the
partial wave decomposition.

\subsection{Standard \textit{K}-Matrix Unitarization}

Following Gupta and collaborators, and subsequent
studies~\cite{Gupta:1949rh,Gupta:1981,Gupta:1993tq,Berger:1991uj,Chanowitz:1999se},
we may reverse the logic behind the definition of the \textit{K}-matrix.  We
interpret the Hermitian \textit{K}-matrix as an incompletely calculated
approximation to the true amplitude, and look for the unitary $S$ or
\textit{T} matrix as a non-perturbative completion of this approximation.

Let us first assume that the scattering matrix is available in
diagonal form.  Given a real eigenamplitude $a_K$~(\ref{eq:aK(a)}) of
the \textit{K} matrix, the corresponding unitarized amplitude $a$ that enters
the \textit{T} matrix is obtained by inverting~(\ref{eq:aK(a)}),
\begin{equation}
\label{eq:a(aK)}
  a = \frac{a_K}{1 - \ii a_K}.
\end{equation}
If the approximation to the scattering matrix \textit{K} is Hermitian but not
available in diagonal form, we can similarly {define} the unitarized
transition matrix \textit{T} as the stereographic projection, by the
formula~(\ref{eq:T(K)}).

The standard \textit{K}-matrix unitarization formalism works on a
perturbative series of the \textit{T} matrix.  Given a $n$-th order
approximation $T_0^{(n)}$ to the \textit{T} matrix, represented by an
eigenamplitude $a_0^{(n)}$, we first have to construct the
corresponding real \textit{K}-matrix amplitude $a_K^{(n)}$
via~\eqref{eq:aK(a)}, 
\begin{align}
\label{eq:a(aK)-pert}
  a_K^{(n)} &= \frac{a_0^{(n)}}{1+\ii a_0^{(n)}}
  = a_0^{(1)} + \Re a_0^{(2)} + \ii(\Im a_0^{(2)} - (a_0^{(1)})^2) +
  \ldots
  \nonumber\\
  &= a_0^{(1)} + \Re a_0^{(2)} + \ldots
\end{align}
where we assume that $a_0^{(1)}$ is real and use the lowest order of
the optical theorem $\Im a_0^{(2)} = (a_0^{(1)})^2$.
At each order, the imaginary parts cancel if the original
perturbation series was correct.  In a second step, we then insert the
truncated perturbation series for $a_K^{(n)}$ into~\eqref{eq:a(aK)},
this time without truncating,
\begin{equation}
\label{eq:a(a0)-K}
  a^{(n)} = \frac{a_0^{(1)} + \Re a_0^{(2)} + \ldots}
  {1 - \ii(a_0^{(1)} + \Re a_0^{(2)} + \ldots)}
\end{equation}
If the exact scattering matrix does admit a perturbative expansion, this
prescription amounts to a partial resummation of the perturbation
series.  In its general form, the construction guarantees that (i) the
computed $S$ matrix is unitary, and (ii) perturbation
theory is reproduced order by order.

For a concrete example, a $2\to 2$ scattering process of scalar
particles with a scalar $s$-channel pole is represented by a $J=0$
partial-wave eigenamplitude
\begin{equation}
  a_{K}^{(0)}(s) = \frac{\lambda}{s - m^2},
\end{equation}
and the unitarized version reads
\begin{equation}
\label{eq:Breit-Wigner}
  a^{(0)}(s) = \frac{\lambda}{s - m^2 - \ii\lambda},
\end{equation}
the Breit-Wigner form of a scalar resonance.  \textit{K}-matrix unitarization, in
this case, therefore implements the Dyson resummation of the resonant
propagator.

Beyond leading order,
given the (non-unitary) perturbative approximation to the transition
matrix \textit{T}, we should reconstruct the corresponding truncated
perturbative expansion of the Hermitian \textit{K} matrix via~(\ref{eq:K(T)})
and insert this back into the unitarization formula~(\ref{eq:T(K)}),
to obtain the corresponding unitarized \textit{T} matrix.  Thus inserting a
$n$th~order approximation of~\eqref{eq:K(T)} into~\eqref{eq:Cayley}
will result in a unitary $S$-matrix to \emph{all} orders.  Conversely,
the $nth$~order expansion of this $S$-matrix will reproduce the original
$nth$~order expression, which is unitary only up to terms of
order~$n+1$.

\subsection{Direct \textit{T}-Matrix Unitarization I: Linear Projection}

While the reconstruction of the unitary $S$ (or \textit{T}) matrix according
to this algorithm is exact within the framework of perturbation
theory, it suffers from the drawback that we have to reconstruct the
self-adjoint \textit{K} matrix as an intermediate step. This is not just
unnecessary, but it may become a significant complication if the
scattering matrix is not available in diagonal form, or if
non-perturbative effects need to be considered.  For practical
purposes, we are rather interested in a means to unitarize an
arbitrary \emph{model} of the scattering matrix, which may or may not
admit a perturbative expansion.

In the following, we therefore present a generalization of the
\textit{K}-matrix prescription that operates on the \textit{T} matrix directly.
Given $a_0$ as a \emph{complex} approximation to an eigenvalue of the
true \textit{T} matrix, we first {define} the unitarized version $a$
by the same geometric construction as before, i.e., connecting the point
$a_0$ with the point $\ii$ by a straight line and determining the
intersection with the Argand circle.  However, we do not attempt to
construct the real amplitude $a_K$.  This results in
\begin{equation}
\label{eq:linear}
  a = \frac{\Re a_0}{1 - \ii a_0^*}
\end{equation}
This formula has the properties that (i) $a$ lies on the Argand
circle, (ii) if $a_0$ is real, it reproduces~(\ref{eq:a(aK)}), and
(iii) if $a_0$ is already on the Argand circle, it is left invariant,
$a=a_0$.  This guarantees the invariance of the correct perturbative
series, up to the resummation of higher orders.  Nevertheless, the
actual expression for~\eqref{eq:linear}, evaluated in perturbation
theory, differs from the standard \textit{K}-matrix
formula~\eqref{eq:a(a0)-K}.  We obtain
\begin{equation}
\label{eq:a(a0)-Kpert}
  a^{(n)} = \frac{a_0^{(1)} + \Re a_0^{(2)} + \ldots}
  {1 - \ii(a_0^{(1)} + \Re a_0^{(2)} - \ii\Im a_0^{(2)} + \ldots)}.
\end{equation}
Due to the truncation of the perturbation series at different stages
of the calculation, higher orders enter in a different way.  We also
note that the standard \textit{K}-matrix formalism, and thus
formula~\eqref{eq:a(a0)-Kpert}, \emph{requires} the existence of a
perturbative series.  By contrast, the direct unitarization
formula~(\ref{eq:linear}) does not rely on a perturbative expansion.
The latter construction is thus applicable to a larger set of models.
In particular, in the case of vector-boson scattering with a light
Higgs that we consider in this paper, the leading term $a_0^{(1)}$ is
suppressed, and thus the original \textit{K}-matrix construction is
ill-behaved.  The modified version~(\ref{eq:linear}) does not suffer
from this problem.

Still, the formula~(\ref{eq:linear}) is not quite satisfactory: if
the imaginary part of $a_0$ becomes larger than $\ii$, the selected
intersection point $a$ appears beyond the fixed point $a=\ii$, on the
complex half-plane opposite to the location of $a_0$.  A consequence
would be that a model amplitude of the form
\begin{equation}
  a_0(s) = \frac{\lambda'}{s - m^2 - i\lambda}
\end{equation}
where $\lambda'>\lambda$, would be transformed into an unitarized
version that revolves \emph{twice} around the Argand circle, splitting
the resonance at $m^2$ into two separate peaks.  Although the original model
is a rather pathological ansatz for a resonance, such a behavior is
clearly undesirable.  To avoid this problem, we may require that (iv)
if $\Im a_0 \geq 1$, the unitarized amplitude $a$ is tied to the fixed
point, i.e., we finally define
\begin{equation}
\label{eq:a-for-Im(a0)>1}
  a = \begin{cases}
    \dfrac{\Re a_0}{1 - \ii a_0^*} &\text{if $\Im a_0 < \ii$},\\
    \ii&\text{otherwise}
    \end{cases}
\end{equation}

We now generalize this prescription to the scattering matrix \textit{T},
starting from a model approximation $T_0$ that is not necessarily
unitary.   We may first restrict ourselves to matrices that are normal
(i.\,e.~$T_0^\dagger T_0 = T_0T_0^\dagger$) and do not have eigenvalues with an
imaginary part larger than $\ii$.  The unitarized transition matrix
then is given by
\begin{equation}
  \label{eq:T(T0)}
  T = \frac{\Re T_0}{\mathbf{1} - \frac{\ii}{2} T_0^\dagger}.
\end{equation}
For non-normal matrices, the operator ordering in the fraction must be
defined.  We obtain two equivalent expressions
\begin{align}
\label{eq:T-nonnormal}
  T &= \frac{1}{\sqrt{\mathbf{1} - \frac{1}{2}\Im T_0}}
      \Re T_0 \frac{1}{\mathbf{1}-\frac{\ii}{2}T_0^\dagger}
      \sqrt{\mathbf{1} - \frac{1}{2}\Im T_0}
 \nonumber\\
    &= \sqrt{\mathbf{1} - \frac{1}{2}\Im T_0}
      \frac{1}{\mathbf{1}-\frac{\ii}{2}T_0^\dagger} \Re T_0 
      \frac{1}{\sqrt{\mathbf{1} - \frac{1}{2}\Im T_0}}\,.
\end{align}
For any matrix $T_0$, the matrix \textit{T} from~(\ref{eq:T-nonnormal})
respects the optical theorem.  If $T_0$ already respects the optical
theorem, we get $T=T_0$.  If $T_0$ represents the correct perturbative
expansion of \textit{T}, truncated at a given order and retaining
non-Hermitian parts, the reconstructed matrix \textit{T} reproduces this
perturbative expansion.  

\begin{figure}
  \begin{center}
    \includegraphics[width=.6\textwidth]{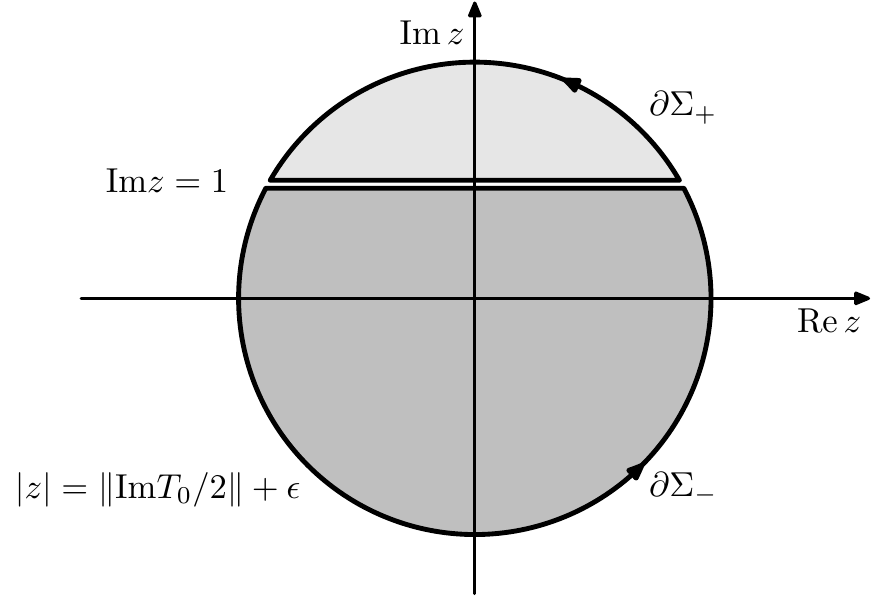}
  \end{center}
  \caption{\label{fig:ImA<1}%
    Integration contours used for projecting on the subspaces
    corresponding to~$\Im T_0/2<1$ and~$\Im T_0/2>1$ for a bounded
    operator~$Im T_0/2$ in~\eqref{eq:T-nonnormal}.}
\end{figure}
Beyond perturbation theory, in order to
extract eigenvalues with imaginary part greater than $\ii$, we may
either diagonalize the matrix and use~\eqref{eq:a-for-Im(a0)>1}, or we
can use projections to make~\eqref{eq:T-nonnormal} well defined.
For this purpose, recall that functions of matrices can be defined by
their power series 
expansion, as long as the radius of convergence exceeds the norm
of the matrix.  More generally, one can use a functional calculus to
associate to a function~$f:D\subseteq\mathbf{C}\to\mathbf{C}$ a
function~$\hat f$ mapping matrices to matrices, such that
\begin{subequations}
\begin{align}
  \widehat{\alpha f + \beta g} &= \alpha \hat f + \beta \hat g \\
  \widehat{f g} &= \hat f \hat g \\
  \widehat{f \circ g} &= \hat f \circ \hat g\,.
\end{align}
\end{subequations}
The Riesz-Dunford functional
calculus~\cite{Riesz:1930,Gelfand:1941,Dunford:1988} 
defines~$\hat f(A)$ by a contour integral encircling the
spectrum~$\sigma(A)$
\begin{equation}
\label{eq:hat-f}
  \hat f(A)
    = \int_{\partial\Sigma:\sigma(A)\subseteq\Sigma}
      \frac{\dd z}{2\pi\ii} \frac{f(z)}{z\mathbf{1}-A}
\end{equation}
using the fact that the resolvent matrix~$1/(z\mathbf{1}-A)$ is well
defined whenever~$z\not\in\sigma(A)$.  Note that this functional
calculus can be used unchanged for all bounded operators on a Hilbert
space.  It can even be extended to certain classes of unbounded
operaters, but the details are not important in the present work,
because we deal with finite dimensional matrices corresponding to
scattering amplitudes with definite angular momentum.
Closely related to this functional calculus~\eqref{eq:hat-f} are the
projections on the invariant subspace of~$A$ corresponding to a
part~$\Sigma\subseteq\sigma(A)$ of the spectrum~\cite{Riesz:1930,Gelfand:1941,Dunford:1988}
\begin{equation}
  P_{A,\Sigma}
    = \int_{\partial\Sigma}\frac{\dd z}{2\pi\ii} \frac{1}{z\mathbf{1}-A}\,.
\end{equation}
In particular we can define projections~$P_{\Im T_0/2, \Sigma_\pm}$ with
\begin{equation}
  \mathbf{1} = P_{\Im  T_0/2, \Sigma_+} + P_{\Im  T_0/2, \Sigma_-}
\end{equation}
using the contours~$\Sigma_\pm$ in fig.~\ref{fig:ImA<1}
to generalize the prescription~\eqref{eq:a-for-Im(a0)>1} for~$\Im T > 2$.

\subsection{Direct \textit{T}-Matrix Unitarization II: Thales Projection}

\begin{figure}[hbt]
\begin{center}
  \includegraphics[width=0.8 \textwidth]{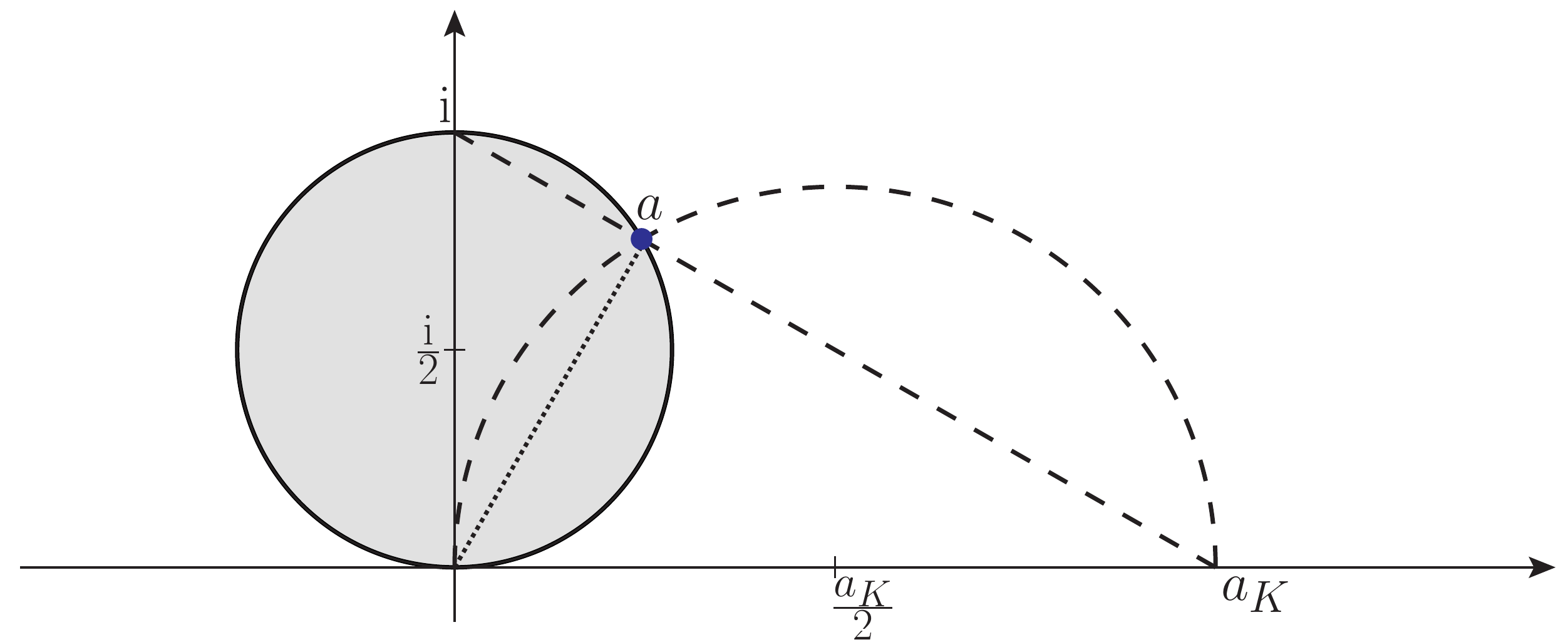}
\end{center}
\caption{Geometrical representation of Thales projection.}
\label{fig:ThalesStereo}
\end{figure}

Elementary geometry (Thales' Theorem) suggests an alternative
construction of the stereographic projection from the real axis to the
unitarity circle, which results
in a different extension to general complex scattering amplitudes.
Fig.~\ref{fig:ThalesStereo} shows that the \textit{K} matrix amplitude
$a_0$ coincides with the endpoint of a half-circle that connects the
lower fixed point $0$ with the unitary amplitude $a$.  Consequently,
given an arbitrary complex amplitude $a$, we define the Thales
projection $a$ as the intersection point of the half-circle that
connects $0$ and $a_0$, with the Argand circle.
The Thales circle is characterized by its intersection $a_K$ with the
real axis, given by
\begin{equation}
	\label{eq:ThalesCircle}
	\left | a-\frac{a_K}{2} \right | = \frac{a_K}{2}.
\end{equation}
Therefore every real amplitude $a_K$ would be projected on the unitary circle
\begin{align}
	a &= \frac{a_K}{1 - \ii a_K}.
\end{align}
In case we start with a complex amplitude $a_0$, we can derive the
transformation to real $a_K$ from the condition, that $a_0$ has to be
on the Thales circle,~\eqref{eq:ThalesCircle}:
\begin{align}
	\frac{1}{a_K} &= \frac{\mathrm{Re}\left (a_0 \right )}{\left |
            a_0 \right |^2}  
		= \mathrm{Re} \left ( \frac{1}{a_0} \right ) \qquad .
\end{align}
We then calculate the transformation for general amplitudes:
\begin{align}
  \label{eq:generalamp}
  a &= \frac{1}{\mathrm{Re} \left ( \frac{1}{a_0} \right ) - \ii}
\end{align}
The corresponding operator equation is
\begin{align}
  \label{eq:generalop}
  T(T_0) &= \frac{1}{\mathrm{Re} \left ( \frac{1}{T_0} \right ) -
    \frac{\ii}{2} \mathds{1}} \qquad .
\end{align}
In appendix~\ref{sec:kmatrixunit} we show that this indeed leads to a
unitary $S$ operator, and that the \textit{T} operation on a $T_0$ operator
is idempotent.

This construction avoids the undesirable behavior for a model amplitude
above the Argand circle; the unitarized version of a single resonance
is again a single resonance.  However it suffers itself from another
undesirable feature: it is not analytic in the vicinity of~$a_0=0$.
Fortunately, this drawback is of little practical importance, because
we are mostly interested in the case where~$a_0\not=0$.

\begin{figure}[hbt]
\begin{center}
  \includegraphics[width=0.8 \textwidth]{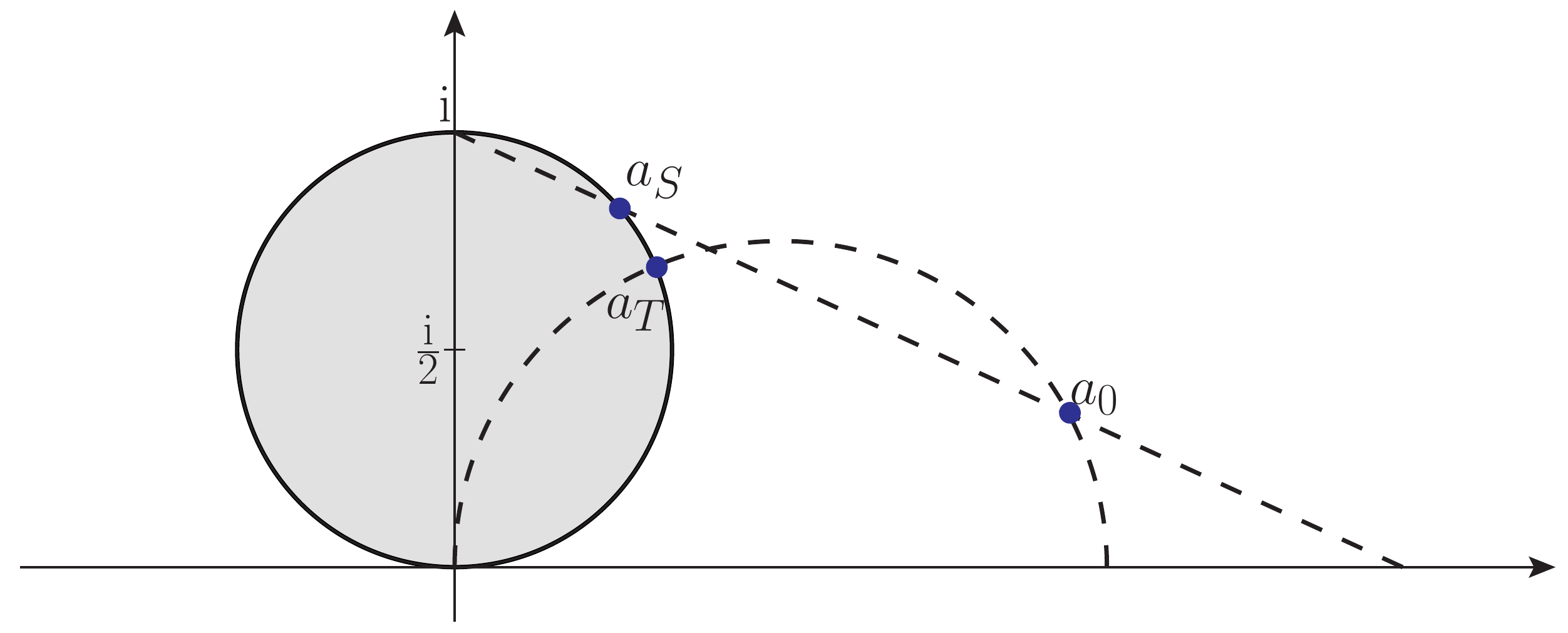}
\end{center}
\caption{Geometrical representation: stereographic projection vs
  Thales projection.} 
\label{fig:ThalesStereo2}
\end{figure}

\subsection{Alternative Unitarization Prescriptions}

The  direct \textit{T}-matrix projection, as described above, allows us
to unitarize any model of the scattering matrix without relying on
perturbation theory or any other details of the processes under
consideration.  It leaves invariant the scattering matrix, if it is
already unitary.  Nevertheless, it is clearly not unique.  Since the
model that we start from does not carry the complete information about
its UV completion, we cannot expect the correct completion to appear
in the unitarized version either.

For an instructive example, consider another parameterless prescription
\begin{subequations}
\label{eq:exp(iL)}
\begin{align}
  S &= \ee^{\ii L} \\
  T &= 2\ee^{\ii L/2} \sin\frac{L}{2}\,,
\end{align}
\end{subequations}
which leads to
\begin{equation}
  L = - \ii \ln S = - \ii \ln (1 + \ii T)\,.
\end{equation}
In a perturbative expansion\footnote{%
  Incidentally, the expansions
  of $L$ and \textit{K} agree in first and second order:
\begin{subequations}
  \begin{align}
    K &= T - \frac{\ii}{2} T^2 - \frac{1}{4} T^3 + \ldots \\
    L &= T - \frac{\ii}{2} T^2 - \frac{1}{3} T^3 + \ldots\,.
  \end{align}
\end{subequations}}
away from the cut starting at $T = - \ii \mathbf{1}$,
the logarithm will be replaced by a
polynomial and $L$ will grow like a power, as the coupling and energy
increases.  In this case, unlike~\eqref{eq:Cayley}, the $S$-matrix
will ``wrap around'' faster and faster, corresponding to a series of
resonances with decreasing distance. 

We do not expect a unitarization prescription to produce additional
structure that is not already present in the original model.  However,
the tower of resonances that appear in~(\ref{eq:exp(iL)}) clearly is
an artefact of the prescription.  From this perspective, the
prescription~\eqref{eq:Cayley}, which for a uniformly growing
amplitude just implies asymptotic saturation and no extra features,
appears to be closer to a minimal and thus natural amendment of the
perturbative prediction.

There are also unitarization prescriptions that rely on reordering a
perturbation series, such as Pad\'e
unitarization~\cite{Pade:1892,Basdevant:1969rw,Basdevant:1969sz}, which has
frequently been applied to vector-boson scattering physics in the
Higgs-less or heavy-Higgs
limit~\cite{Dobado:1989gr,Dobado:1990jy,Dobado:1990am,Dicus:1990ew,Gupta:1993tq}.
This method reproduces certain exactly solvable
models~\cite{Willenbrock:1990dq}.  Unitarization
prescriptions of this kind tend to generate resonances (poles)  at higher energy that are not present in the original EFT.  Similar
effects are observed when applying the
inverse-amplitude~\cite{Truong:1988zp,Truong:1991gv,Dobado:1996ps,Dobado:1999xb,Espriu:2014jya} or
N/D unitarization prescriptions~\cite{Truong:1991ab,Truong:1991gv}.
This may be useful if the amplitude  in
  the correct UV completion actually contains those
resonances (as in pion-pion scattering).

Other approaches explicitly apply a form-factor suppression to
amplitudes that nominally violate unitarity
constraints~\cite{Zeppenfeld:1987ip,Baur:1987mt,Figy:2004pt,Hankele:2006ma}.
Such a suppression indicates new physics, e.g., mixing with nearby
resonances or additional open channels which dissipate the scattering
into multi-particle final states.  This is a possible scenario for
high-energy electroweak interactions, but it is not a prediction of
unitarity~\cite{Degrande:2012wf}.  The form factors depend on
additional parameters.  In order to implement such a behavior, one
would describe the new physics explicitly.

\subsection{Unitarization as a Framework}

In the example computations below, we explicitly apply direct $T$-matrix projection
to quasi-elastic scattering in the Goldstone limit, at
tree level, and extend the results to full scattering amplitudes and
cross sections.  At this level, it coincides with the $K$-matrix
prescription for elastic scattering, analogously extended.  However,
we emphasize that the direct $T$-matrix method is of generic nature,
since it allows us to unitarize any model for any class of processes,
limited just by calculability of the actual expressions.

In particular, we may consider loop-corrected EFT
  amplitudes as the starting point for unitarization.  These
  amplitudes provide an imaginary part, which is correctly treated by
  the prescription and accounted for in the resummation.  The
  resummation corrects the perturbative amplitude by terms which are
  formally of higher order, but become relevant and restore unitarity
  once the growing amplitude enters the strongly interacting regime.
  Likewise, we may choose to insert an amplitude that has already been
  unitarized by any of the above mentioned unitarization
  prescriptions.  In that case, the $T$-matrix prescription will leave
  the amplitude unchanged.  Furthermore, it is possible to apply the
  method to all polarization components of vector bosons, without
  recourse to the Goldstone limit,
  and to properly incorporate $2\to n$ processes.

In short,
  the prescription that we propose serves as a
  \emph{framework} which we can implement not just for the extrapolation of
  the tree-level EFT result 
  (see below), but to any more sophisticated description of VBS,
  or electroweak processes in general.

In the following section, we evaluate
 the direct \textit{T}-matrix
projection for to the minimal
EFT with anomalous couplings, as a simple application.  We do not expect a UV complete model to
emerge.  The implemented asymptotics is minimal, interpolating the
low-energy EFT with high-energy unitarity saturation for any parameter
set different from the SM.  We propose to take this as a class of
reference models.  As soon as experiment will allow us to inspect the
high-energy behavior in more detail, we should introduce specific
extensions, such as new resonances or other kinds of new physics,
similarly applying $T$-matrix projection where necessary. Such refined models, which could
be the result of one of the more predictive scheme as discussed above,
can then be compared to the reference model in the analysis of actual
data.

\section{Unitary Description of Electroweak Interactions}

\subsection{Unitarity for electroweak scattering amplitudes}

In the current paper, we are interested in a model-independent
bottom-up approach to VBS processes.  The Higgs-induced cross-section
suppression makes VBS a prime candidate for looking at anomalous
effects.  Furthermore, there are possible extensions of the SM which provide
large (tree-level) contributions exclusively to the quartic couplings,
via resonance exchange in s- and t-channels, but only minor
contributions to dimension-six operators in the EFT~\cite{Arzt:1994gp}.  

Results from
analyzing VBS data should be combined with all kinds of different
measurements, many of which remain well defined in the EFT.  However,
the EFT breakdown within the accessible region inevitably introduces a
model dependence.  We should set up the phenomenological description
in such a way that this model dependence is kept under control.

The fundamental process in question is a quasi-elastic $2\to 2$
scattering process of Goldstone bosons.  The unitarity requirement
takes a particularly simple form, since we can employ angular-momentum
and isospin symmetry to completely diagonalize the scattering
process.  The eigenamplitudes are just scalar functions of $s$ which
must satisfy the Argand-circle condition~(\ref{eq:Argand}) as long as
no inelastic channels appear.

\begin{figure}[hbt]
\begin{center}
  \includegraphics[scale=0.3]{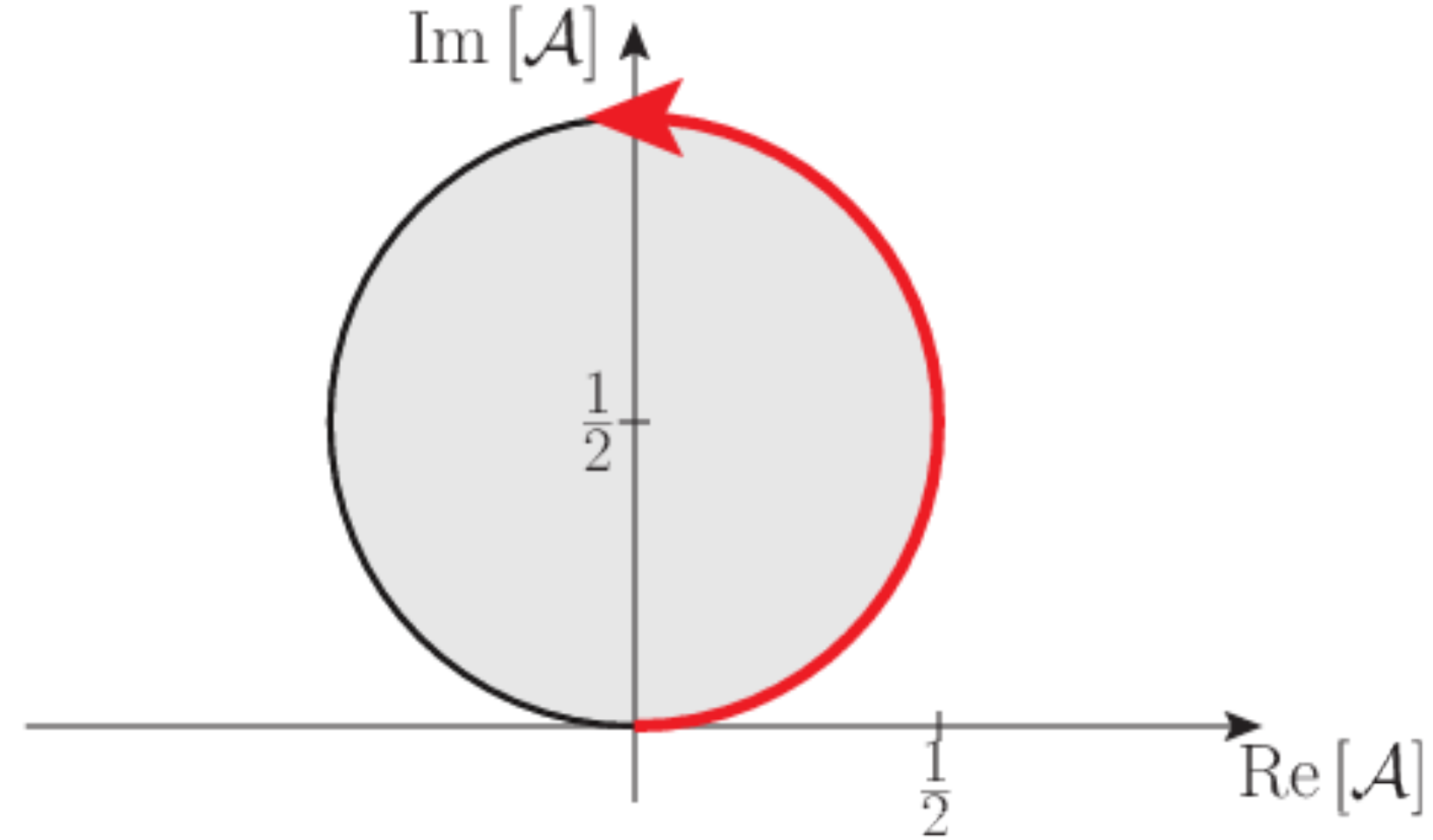}
  \quad
  \includegraphics[scale=0.3]{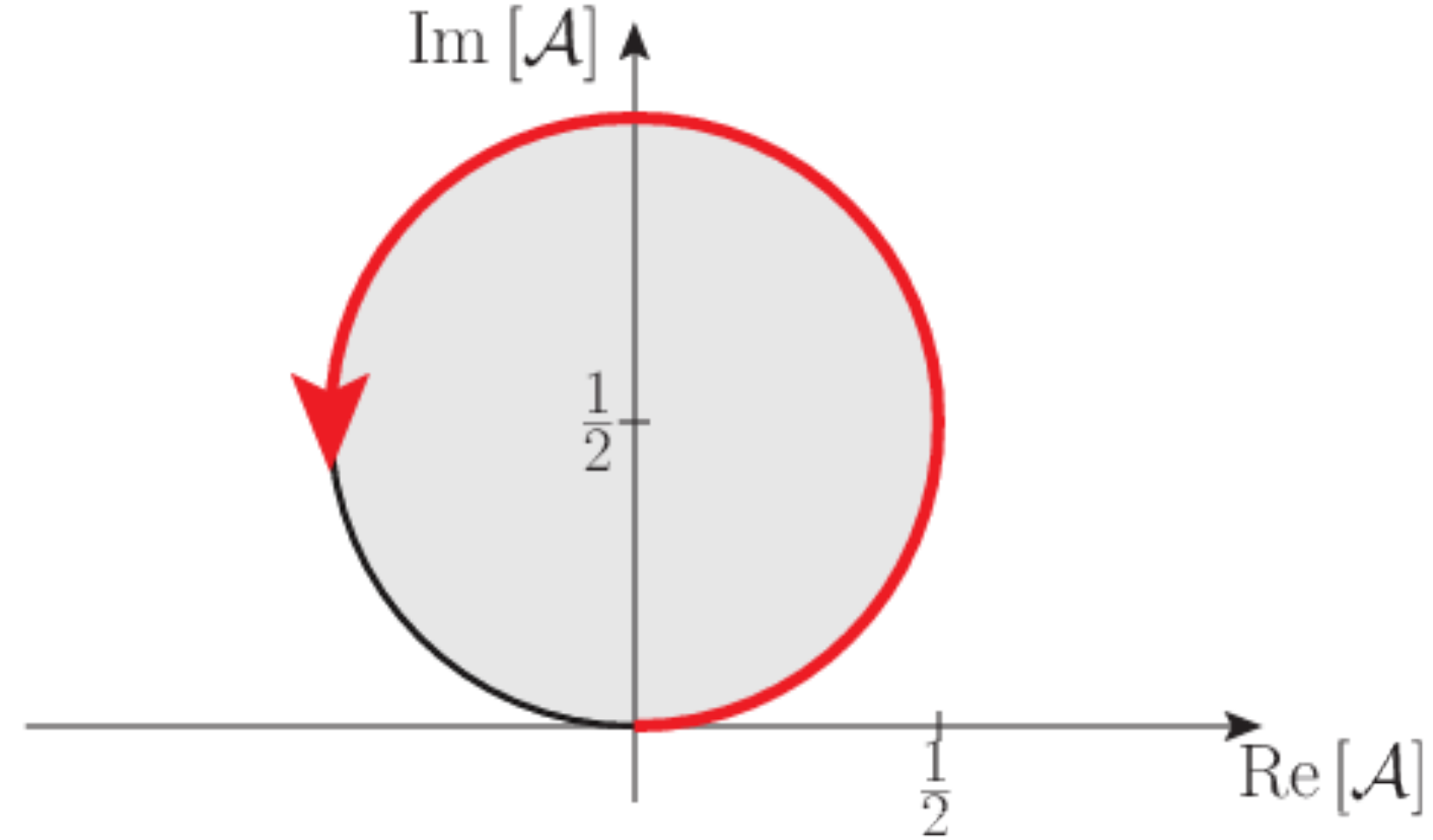}
  \quad
  \includegraphics[scale=0.3]{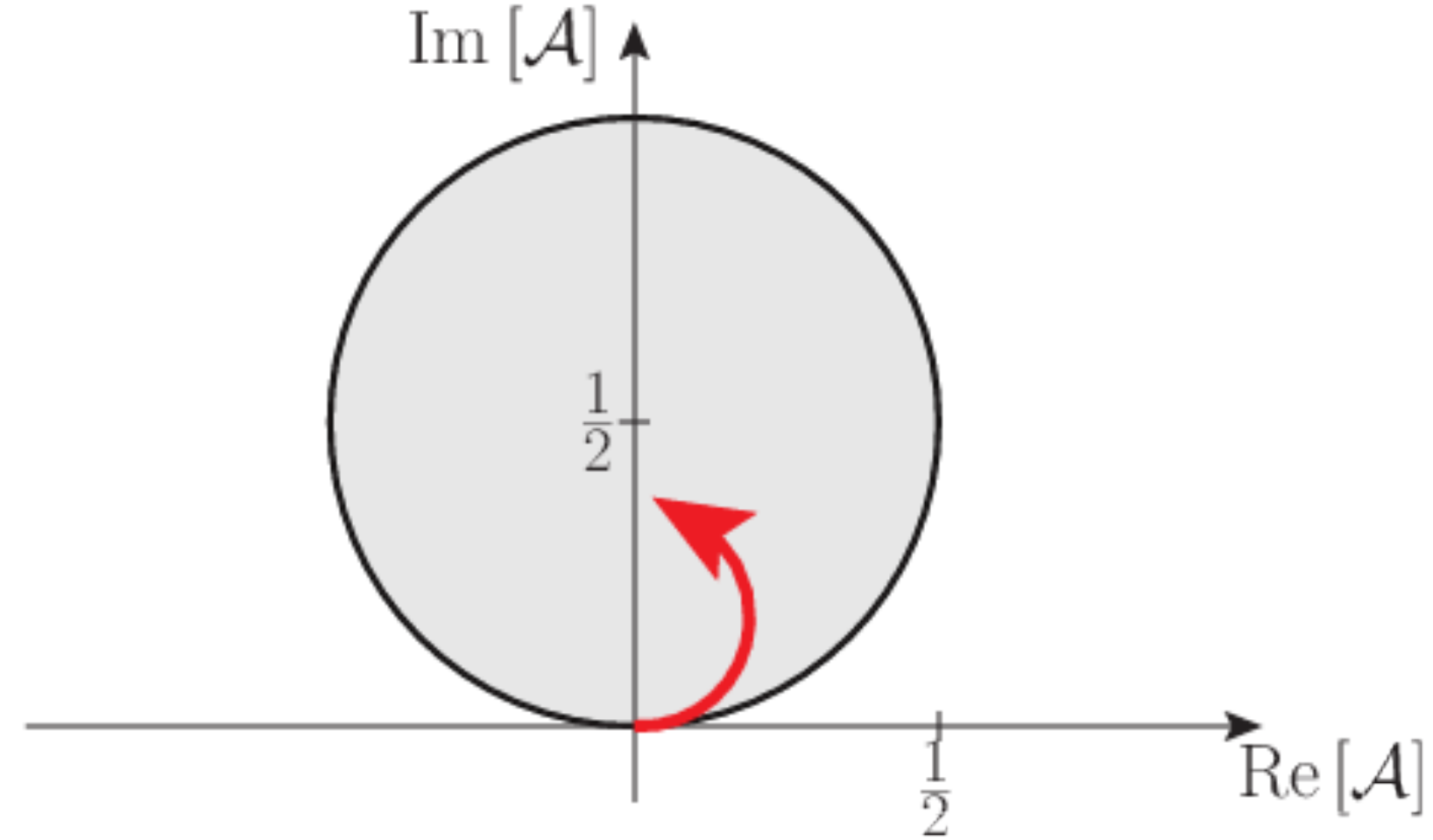}
\end{center}
\caption{The Argand-circle condition for a scattering amplitude.}
\label{fig:argand}
\end{figure}

In reality, we should face one of the following situations:
\begin{enumerate}
\item
  the amplitude stays in the perturbative regime, close to zero, and
  the imaginary part is small compared to the real part.  This is the
  SM case.
\item
  the amplitude rises beyond this level.  Then, it will develop an
  imaginary part, and we are in a strongly interacting regime.  This
  happens if there is any dimension-eight operator with a noticeable
  coefficient.
\item
  the amplitude approaches the maximimum absolute value,
  asymptotically (Fig.~\ref{fig:argand}, left).
\item
  the amplitude turns over.  This is a resonance
  (Fig.~\ref{fig:argand}, center).
\item
  new inelastic channels open and
  absorb part of the total cross section (Fig.~\ref{fig:argand},
  right).  This amounts to an increase in
  the amplitude that is halted by
  an effective form-factor suppression.
  The extra channels, typically resulting in multiple vector boson
  production, should then be observable~\cite{Chanowitz:1984ne,Barger:1991jn}.
\end{enumerate}
For a prediction, we have to make a choice among these possibilities.
There is no case where the amplitude (in the ideal case of pure
Goldstone scattering) leaves the Argand circle, so the naive EFT
result is no option.

In general, apart from the exact SM case we are necessarily in a
non-perturbative regime.  In line with the discussion in the preceding
section, we propose to take case 3.) as a reference model for the
high-energy behavior, correctly matched to the low-energy EFT.  This
idea is realized by the parameter-free direct \textit{T}-matrix unitarization
prescription, as an extension of the \textit{K}-matrix unitarization
formalism as described in the preceding Section.  In the high-energy range, the results saturate the unitarity
bound.  We thus obtain an approximate \emph{upper bound} for the set
of possible amplitudes that match to a given EFT.

We recall that the \textit{T}-matrix prescription is not a viable UV
completion of the EFT, but should be understood as a safeguard against
computing unphysical contributions beyond the unitarity limit.  In the
case at hand, the unitarization changes the interpretation of EFT
operator coefficients.  While they formally remain the parameters of a
low-energy Taylor expansion of the cross section, they effectively
take the role of threshold parameters that indicate the point of
energy where the differential cross section deviates from the SM
prediction and enters a strongly interacting regime.  This threshold
region is the energy range to which the experimental analysis will be
most sensitive.  In a context where the EFT applies, they keep their
relation to the full set of operator coefficients that may be
determined by a global fit to experimental results.  The high-energy
range where actual model dependence becomes important, is
asymptotically suppressed in the same way as the SM prediction and has
a minor impact on observed experimental data, as one would expect.

A complete description of the processes in question
should aim at a more detailed understanding. 
Such a refinement typically relies on more specific assumptions or introduces new parameters.
However, the first measurements of VBS will not be very sensitive to details
beyond threshold, so simulations based on a simple unitarization
prescription in the EFT context will at least allow us to quantify the
level of agreement (or disagreement) of data with the SM.

\subsection{Model and Calculation: Amplitudes}

There are various refinements that we must apply to the idealized
model of the preceding section.  (i) We have to translate
Goldstone-boson interactions to interactions of vector bosons.  This
introduces the explicit $SU(2)_C$ breaking associated with
hypercharge.  (ii) Further low-energy corrections are
caused by the mixing of transversal and longitudinal (effectively
scalar) polarization components.  This mixes spin and orbital angular
momentum and spoils the simplicity of the partial-wave expansion.
(iii) The vector bosons are off-shell, in particular in the initial
state.  (iv) In the forward scattering direction, massless photon
exchange becomes relevant, cut off only by the off-shellness of the
vector bosons in the initial state, and thus a significant correction.

We approach this situation by the following algorithm.  First, we analytically
unitarize the scattering amplitudes in the high-energy limit where the
symmetries are exact, and the unitarity-violating terms occur exclusively in
Goldstone-boson scattering.  In particular, we can make use of
custodial, i.e., weak isospin symmetry and thereby reduce the number
of independent amplitudes.

In contrast to the no-Higgs case which has been discussed extensively
in the literature, in the presence of a light Higgs boson, the SM
contribution to Goldstone scattering is asymptotically suppressed
proportional to $m_H^2/(4\pi v)^2$, compared to a value which would
saturate the unitarity bounds.  In practice
(cf.~Fig.~\ref{fig:W+W+bare}), the suppression in the differential
cross section is an order of magnitude, and the formally subleading
transversal degrees of freedom dominate the observable cross section.
Thus, we take the SM contribution to the
Goldstone-scattering amplitudes as zero.  Nonzero
contributions are induced by the anomalous operators, which thus
become leading. (The spin eigenamplitudes $\mathcal{A}$ have to be
normalized by $a_0=\frac{\mathcal{A}}{32 \pi}$.)
\begin{align}
	\mathcal{A}(w^+w^+ \rightarrow w^+w^+)
        =\ & \frac{1}{4} F_{S,0}(2s^2+t^2+u^2) 
		+\frac{1}{2} F_{S,1} (t^2+u^2)\nonumber\\
		&- \left (F_{HD}^2\frac{v^2}{4}+F_{HD} \right )
		\left( \frac{t^2}{t-m_H^2} +\frac{u^2}{u-m_H^2} \right)\\
	\mathcal{A}(w^+z \rightarrow w^+z) =\ &
        \frac{1}{4} F_{S,0}(s^2+u^2) 
		+\frac{1}{2} F_{S,1} t^2
		- \left (F_{HD}^2\frac{v^2}{4}+F_{HD} \right )
		\frac{t^2}{t-m_H^2} \\
	\mathcal{A}(w^+w^- \rightarrow w^+w^-) 
        =\ &  \frac{1}{4} F_{S,0}(s^2+t^2+2u^2) 
		+\frac{1}{2} F_{S,1} (s^2+t^2)\nonumber\\
		&- \left (F_{HD}^2\frac{v^2}{4}+F_{HD} \right )
		\left( \frac{s^2}{s-m_H^2}+ \frac{t^2}{t-m_H^2} \right) \\
	\mathcal{A}(w^+w^- \rightarrow zz) =\ &
        \frac{1}{4} F_{S,0}(t^2+u^2) 
		+\frac{1}{2} F_{S,1} s^2
		- \left (F_{HD}^2\frac{v^2}{4}+F_{HD} \right ) \frac{s^2}{s-m_H^2} \\
	\mathcal{A}(zz \rightarrow zz) =\ &
        \frac{1}{2}\left (F_{S,0}+F_{S,1}\right ) 
		(s^2+t^2+u^2) \nonumber\\
		&- \left (F_{HD}^2\frac{v^2}{4}+F_{HD} \right )
		\left( \frac{s^2}{s-m_H^2}+ \frac{t^2}{t-m_H^2} +\frac{u^2}{u-m_H^2} \right)
\end{align}
Note that $s > m_H^2$ for the observed Higgs boson, so there are
actually no poles in the physical region.

In the presence of the Higgs boson, there are
  also amplitudes that involve external Higgs bosons.  In terms of
  custodial $SU(2)$, the Higgs is a singlet, and there are additional
  independent amplitudes that involve either two or four Higgs bosons.
  Realistically, the only experimentally accessible channels are
  $w^+w^-\to hh$ and $zz\to hh$.  These channels provide an independent
  set of observables, and they should be studied in
  the context of a larger set of operators.  This is beyond the
  scope of the present paper.

Higgs-pair channels also contribute implicitly to the
  unitarization condition via backscattering into Goldstones, and thus
  physical vector bosons.  Among the set of operators that we consider
  in the above example, only the dimension-six term $\LL_{HD}$
  provides a Higgs-pair contribution.  The Standard-Model
  contribution, given the approximations, is equivalent to zero.  In
  the following, we neglect this extra contribution for simplicity.
  Adding it, we would have to apply a further diagonalization of
  eigenamplitude in the isospin-zero channel and unitarize the
  resulting independent eigenamplitudes.  Transforming back to the
  physical basis, the effective $F_{HD}$ term is slightly more
  suppressed.  However, as we will see in
  the final plots, the $\LL_{HD}$ term is of minor importance anyway.
  For the purposes of the example, we therefore keep the formulae
  simple and omit this contribution.  In a more complete treatment
  that applies the unitarization framework to the full set of
  operators and also unitarizes transversal contributions from these
  sources, it can be properly incorporated.
  
Since all operators are $SU(2)_C$-symmetric, we can apply isospin
symmetry and crossing symmetry and express all amplitudes in terms of
a single master amplitude $A(s,t,u)$~\cite{Lee:1977yc},
\begin{align}
\label{eq:A(ww->zz)}
	\mathcal{A}(w^+w^- \rightarrow zz) &= A(s,t,u), \\
	\mathcal{A}(zz \rightarrow zz) &= A(s,t,u) + A(t,s,u) + A(u,s,t), \\
\label{eq:A(ww->ww')}
	\mathcal{A}(w^+w^- \rightarrow w^+w^-) &= A(s,t,u) + A(t,s,u), \\
\label{eq:A(wz->wz)}
	\mathcal{A}(w^+z \rightarrow w^+z) &= A(t,s,u), \\
\label{eq:A(ww->ww)}
	\mathcal{A}(w^+w^+ \rightarrow w^+w^+) &= A(t,s,u) + A(u,s,t),
\end{align}
and construct the isospin eigenamplitudes $\mathcal{A}_I$,
\begin{align}
	\mathcal{A}_2&= A(t,s,u) + A(u,s,t), \\
	\mathcal{A}_1&= A(t,s,u) - A(u,s,t), \\
	\mathcal{A}_0&= 3A(s,t,u)+ A(t,s,u) + A(u,s,t).
\end{align}
After partial wave decomposition ($t = -s/2 (1-\cos\Theta)$)
\begin{align}
	A_{I\ell}(s) = \int_{-s}^0 \frac{dt}{s}
	\mathcal{A}_I(s,t,u)P_\ell\left(\cos\Theta\right).
\end{align}
we obtain the isospin-spin eigenamplitudes:
\begin{subequations}
\label{eq:isospin-spin}
\begin{align}
	\mathcal{A}_{00} = & \frac{1}{6} \left (7 F_{S,0} + 11F_{S,1}
        \right )s^2 \nonumber\\ 
		&-\left (F_{HD}^2\frac{v^2}{4}+F_{HD} \right )
		 \left ( \frac{3s^2}{s-m_H^2} + 2 \mathcal{S}_0(s)
                 \right), \\ 
	\mathcal{A}_{02} = & \frac{1}{30} \left (2 F_{S,0} + F_{S,1} \right )s^2 
		-\left (F_{HD}^2\frac{v^2}{4}+F_{HD} \right ) 2
                \mathcal{S}_2(s), \\ 
	\mathcal{A}_{11} = & \frac{1}{12} \left ( F_{S,0} - 2 F_{S,1} \right )s^2 
		-\left (F_{HD}^2\frac{v^2}{4}+F_{HD} \right ) 2
                \mathcal{S}_1(s), \\ 
	\mathcal{A}_{13} = & -\left (F_{HD}^2\frac{v^2}{4}+F_{HD}
        \right ) 2 \mathcal{S}_3(s), \\ 
	\mathcal{A}_{20} = & \frac{1}{3} \left (2 F_{S,0} + F_{S,1} \right )s^2 
		-\left (F_{HD}^2\frac{v^2}{4}+F_{HD} \right ) 2
                \mathcal{S}_0(s), \\ 
	\mathcal{A}_{22} = & \frac{1}{60} \left (F_{S,0} + 2 F_{S,1} \right )s^2 
		-\left (F_{HD}^2\frac{v^2}{4}+F_{HD} \right ) 2 \mathcal{S}_2(s),
\end{align}
\end{subequations}
using following abbreviations from \cite{Alboteanu:2008my} 
\begin{subequations}
\begin{align}
	\mathcal{S}_{0} &= m_H^2+\frac{m_H^4}{s}\log \left (\frac{m_H^2}{s + m_H^2} \right )
		-\frac{s}{2}, \\ 
	\mathcal{S}_{1} &= 2 \frac{m_H^2}{s}
		+\frac{m_H^4}{s^2} \left( 2 m_H^2 +s \right )
		\log \left (\frac{m_H^2}{s + m_H^2} \right )
		+\frac{s}{6}, \\ 
			\mathcal{S}_{2} &=  \frac{m_H^4}{s^2} 
			\left (6 m_H^2 + 3s \right )
		+\frac{m_H^4}{s^3} \left( 6 m_H^4 +6m_H^2 s +s^2 \right )
		\log \left (\frac{m_H^2}{s + m_H^2} \right ).
\end{align}
\end{subequations}
Expressed in terms of the isospin-spin eigenstates, the Goldstone
scattering matrix becomes diagonal.

It is now straightforward to apply the \textit{T}-matrix unitarization scheme
(equivalent to the \textit{K}-matrix scheme at this order)
to the diagonal isospin-spin eigenamplitudes.  For each $I\ell$
combination, the \textit{T}-matrix unitarized amplitude is given by,
cf.~\eqref{eq:generalamp},
\begin{align}
	\hat{\mathcal{A}}_{I\ell}(s)= \frac{1}
		{\text{Re} \left( \frac{1}{\mathcal{A}_{I\ell}(s)} \right)
		-\frac{\ii}{32\pi} }.
\end{align}
We split off the original amplitude $\mathcal{A}_{I\ell}$ that
corresponds to the naive EFT and obtain the unitarization correction
as
\begin{align}
	\Delta
        \mathcal{A}_{I\ell}=\hat{\mathcal{A}}_{I\ell}-\mathcal{A}_{I\ell}. 
  \label{eq:amplitudeCT}
\end{align}
Given this set of corrections, we dress the eigenamplitude corrections
by the appropriate Legendre polynomials and revert the basis from
isospin eigenstates to $w^+,z,w^-$, so we arrive at counterterms for
the individual Goldstone scattering channels:
\begin{align}
  \Delta \amp(w^+w^+\rightarrow w^+w^+ ) 
  &=  \Delta \amp_{20}(s)-10\Delta \amp_{22}(s)
  \notag\\
  &+ 15 \Delta \amp_{22}(s) \frac{t^2+u^2}{s^2},
\label{eq:Delta-A(ww->ww)}
  \\
  \Delta \amp(w^+w^-\rightarrow zz ) 
  &= \frac{1}{3} \left ( \Delta \amp_{00}(s)-\Delta \amp_{20}(s)
  \right )-\frac{10}{3}\left ( \Delta \amp_{02}(s)-\Delta \amp_{22}(s)
  \right )
  \notag\\
  &+ 5 \left (\Delta \amp_{02}(s)-\Delta \amp_{22}(s) \right)
  \frac{t^2+u^2}{s^2},
\label{eq:Delta-A(ww->zz)}
  \\
  \Delta \amp(w^+z\rightarrow w^+z ) 
  &= \frac{1}{2}\Delta \amp_{20}(s)-5\Delta \amp_{22}(s) 
  \notag\\
  &+ \left ( -\frac{3}{2}\Delta \amp_{11}(s)+\frac{15}{2}\Delta
    \amp_{22}(s) \right )\frac{t^2}{s^2}
\label{eq:Delta-A(wz->wz)}
  \notag\\
  &+  \left (\frac{3}{2}\Delta \amp_{11}(s)+\frac{15}{2}\Delta
    \amp_{22}(s) \right )\frac{u^2}{s^2},
  \\	
  \Delta \amp(w^+w^-\rightarrow w^+w^- ) 
  &= \frac{1}{6} \left ( 2\Delta \amp_{00}(s)+\Delta \amp_{20}(s)
  \right )-\frac{5}{3}\left (2 \Delta \amp_{02}(s)+\Delta \amp_{22}(s)
  \right )
  \notag\\
  &+ \left ( 5\Delta \amp_{02}(s)-\frac{3}{2}\Delta
    \amp_{11}(s)+\frac{5}{2} \Delta \amp_{22}(s) \right )\frac{t^2}{s^2}
  \notag\\
  &+ \left ( 5\Delta \amp_{02}(s)+\frac{3}{2}\Delta
    \amp_{11}(s)+\frac{5}{2} \Delta \amp_{22}(s) \right )
  \frac{u^2}{s^2},
\label{eq:Delta-A(ww->ww')}
  \\
  \Delta \amp(zz\rightarrow zz ) 
  &= \frac{1}{3} \left ( \Delta \amp_{00}(s)+2\Delta \amp_{20}(s)
  \right )-\frac{10}{3}\left ( \Delta \amp_{02}(s)+2\Delta
    \amp_{22}(s) \right )
  \notag\\
  &+ 5 \left (\Delta \amp_{02}(s)+2\Delta \amp_{22}(s) \right )
  \frac{t^2+u^2}{s^2}.
\label{eq:Delta-A(zz->zz)}
\end{align}
Since there are no branch cuts in lowest order, crossing symmetry
implies that the amplitude 
$\mathcal{A}(w^+z \rightarrow w^+z)$ in \eqref{eq:A(wz->wz)}
can be obtained from
$\mathcal{A}(w^+w^- \rightarrow zz)$ in \eqref{eq:A(ww->zz)}
by an exchange of~$s$ with~$t$ and
$\mathcal{A}(w^+w^+ \rightarrow w^+w^+)$ in \eqref{eq:A(ww->ww)}
from 
$\mathcal{A}(w^+w^- \rightarrow w^+w^-)$ in \eqref{eq:A(ww->ww')}
by an exchange of~$s$ with~$u$ (using $A(s,t,u)=A(s,u,t)$).
In the presence of branch cuts, however, there are contributions
like resonance poles on the unphysical Riemann sheet
that must only be non-zero if the Mandelstam variable is above the
threshold~$M^2_{\text{thr.}}$ of the corresponding branch cut.  The
resulting factors~$\Theta(s-M^2_{\text{thr.}})$ and additional terms
proportional to~$\Theta(t-M^2_{\text{thr.}})$
and~$\Theta(u-M^2_{\text{thr.}})$ have been suppressed
in~(\ref{eq:Delta-A(ww->ww)}-\ref{eq:Delta-A(zz->zz)}), with the
understanding that these formulae will only be used in the
case~$s>0\land t<0\land u<0$.

As a toy example for this phenomenon, consider a
unitarity-violating amplitude
\begin{equation}
  \mathcal{A}(s) = \alpha \frac{s}{v^2}
\end{equation}
resulting from a local dimension-six operator, that can be unitarized by
replacing the local operator by a resonance exchange
\begin{equation}
\label{eq:A_R}
  \mathcal{A}_R(s) =
     - \frac{s}{s - M^2 + \ii \Gamma M \Theta(s-M^2_{\text{thr.}})}
\end{equation}
with~$M^2=v^2/\alpha$, so that $\mathcal{A}(s)$ and~$\mathcal{A}_R(s)$ agree in the
region~$s\ll M^2$.  Since the amplitude must not have a pole with
a non-vanishing imaginary part on the physical Riemann sheet, any such
pole is located on the second sheet, which can be reached via a branch
cut corresponding to an open decay channel of the
resonance~\cite{Analytic-S-Matrix}.
Therefore the pole of the crossed amplitude in the $t$-channel
\begin{equation}
\label{eq:A_R-crossed}
  \bar{\mathcal{A}}_R(t) =
     - \frac{t}{t - M^2 + \ii \Gamma M \Theta(t-M^2_{\text{thr.}})}
\end{equation}
must not have the imaginary part, because there is no branch cut
for~$t<0$ providing access to the second sheet.
In~\eqref{eq:A_R} and~\eqref{eq:A_R-crossed}, this is
expressed by the $\Theta$-distributions multiplying the widths.

\begin{figure}
  \begin{center}
    \includegraphics[width=.6\textwidth]{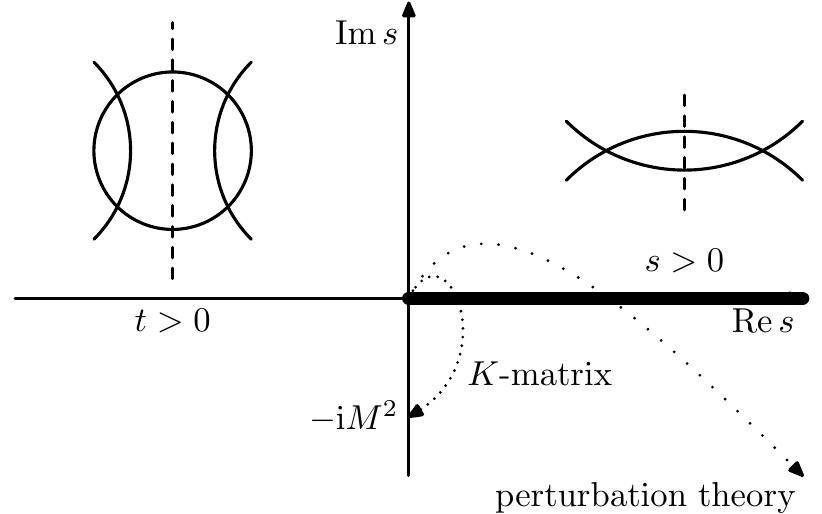}
  \end{center}
  \caption{\label{fig:cuts}%
    Sketch of the analytic structure of resummed scattering
    amplitudes.  As illustrated by the Feynman diagrams, the
    lefthanded (i.\,e.~$t>0$ or~$s<0$) cut will in general
    belong to a different channel (i.\,e.~combination of quantum
    numbers) than the righthanded cut. The diagrams also illustrate
    that the lefthanded cut can appear in a different order of the
    perturbative expansion than the righthanded
    cut. Following~\cite{Willenbrock:1990dq}, we also compare the
    location of the pole in the $K$-matrix scattering
    amplitude~\eqref{eq:A_K} to the perturbative result. Note that
    the passage through the region with $\Im s>0$ has been
    exaggerated for illustrative purposes
    (cf.~\cite{Analytic-S-Matrix}).  In reality, the pole
    immediately crosses over to the second sheet.}
\end{figure}
The analogous analytical structure is found by rewriting the
$K$-matrix unitarized amplitude
\begin{equation}
  \mathcal{A}_K(s) = \frac{\alpha s / v^2}{1 - \ii \alpha s / v^2}
\end{equation}
as
\begin{equation}
\label{eq:A_K}
  \mathcal{A}_K(s) = \frac{\ii s}{s + \ii M^2}
\end{equation}
to make the existence of a similar pole off the real axis explicit.
Again, such a pole must not be located
on the physical Riemann sheet and we may
replace~$\mathcal{A}(s)$ by~$\mathcal{A}_K(s)$ only
for~$s>0$, where a nearby branch cut can act as a portal to the
second sheet, as shown in Fig.~\ref{fig:cuts}.
Note that this prescription manifestly unitarizes
all partial wave amplitudes for~$s\to\infty$, even though the
amplitude as a function of~$s$ and~$t$ appears to
rise for $t\to-s$ for each fixed~$s$. The consistency of the
analytical structure illustrated in Fig.~\ref{fig:cuts} can also be
seen in a perturbative example: in~\cite{Willenbrock:1990dq} it has
been shown explicitely for the example of the $\mathrm{O}(2N)$ model
with large $N$, that the $K$-matrix prescription~\eqref{eq:A_K}
reproduces the exact amplitude when resummed to all orders.

\subsection{Complete Electroweak Processes}
\label{sec:completeEW}

So far, we have only considered Goldstone-scattering amplitudes, which
represent longitudinal vector bosons at asymptotically high energy.
The result of the unitarization procedure is a set of correction terms
that depend on $s,t,u$.  We would like to use the expressions in a
calculation of vector-boson scattering amplitudes at finite energy.
To achieve this, we note that by construction, the counterterms have a
$t$ and $u$ dependence that is equivalent to the anomalous quartic
terms that we started with.  We can therefore unambiguously distribute
the new contributions among the two different gauge-invariant
interaction operators (counterterms) $\LL_{S,0}$ and $\LL_{S,1}$ that
are already present.  The algorithm follows precisely the derivation
in~\cite{Alboteanu:2008my}.

In the result, all three parameters $F_{DM},F_{S,0},F_{S,1}$ enter
both of the counterterm prefactors, respectively.  The unitarization
procedure effectively modifies and mixes the EFT operator coefficients in a
nonlinear way.

We finally switch back from covariant gauge to unitarity gauge and
obtain Feynman rules for physical vector bosons.  Inserting external
momenta and polarization vectors (or fermionic currents), the
asymptotic amplitude expressions receive finite low-energy corrections
that are related to the $W$ and $Z$ masses, some of them breaking the
custodial symmetry.

In the context of complete scattering amplitudes, the new Feynman
rules for quartic gauge-boson couplings are evaluated off-shell.  We
have to define a prescription that determines the energy value in the
operator coefficient.  Relying on the assumption that the effective
vertices are evaluated for an approximately on-shell $2\to 2$
scattering kinematics, we define the energy value as the square root
of the initial- or final-state invariant mass, i.e., the $\sqrt{\hat
  s}$ value for the $VV$ system, represented by their decay products.
This completes the algorithm.

Before we turn to concrete results, we should review the
assumptions and approximations on which the algorithm is based.  First
of all, we started from the linear Higgs EFT as the low-energy
approximation and assume the absence of new states (resonances) within
the accessible energy range.  We have to accept the fact that the
model enters a strongly interacting regime, so beyond the threshold
where the corrections start to play a role, the prediction
becomes a rather uncertain estimate, controlled just by the
unitarity requirement.  However, the unitarized cross section
asymptotically falls off, so the energy range beyond this threshold is
again suppressed in the event sample.  Finally, the unitarization
corrections are strictly valid only in the high-energy
limit and for on-shell longitudinal vector bosons in the kinematical configuration
of quasi-elastic scattering.  We thus have to require that
these conditions are approximately met, typically by imposing VBF cuts in the
analysis.  

Anomalous interactions of
transversal vector bosons would also require unitarization.  They
can be incorporated, analyzing 
them in the high-energy limit where they are decoupled from
Goldstone bosons and then applying the same scheme to the corrected
amplitudes.  However, we do not attempt this explicitly in the
present paper.

These constraints imply, in particular, that the results can
\emph{not} be applied to the analogous process of triple vector boson
production.  The SM, and any underlying UV completion would allow us to determine the correct
analytic structure and relations between processes that are related by
crossing external particles between the incoming and outgoing states.
However, the unitarization corrections in the
present model apply only to s-channel kinematics in $2\to 2$
scattering and must not be used
for the kinematical configuration of triple-boson production
where the initial vector boson is far off-shell.  
We may compare this situation to the resummation
of the propagator of an unstable particle which may also occur in the
t-channel.  In the context of SM gauge invariance, it is necessary to
include extra diagrams in the unitarized
result~\cite{Argyres:1995ym}.  In the present case where the complete
theory is not even known, the corresponding ambiguity is an indication of
the unavoidable model dependence of the unitarization procedure.

\subsection{Numerical Results: On-Shell}

We have implemented the Feynman rules that correspond to the
energy-dependent counter\-term operators, as described in the preceding
Section, in the Monte-Carlo event generator
WHIZARD~\cite{Kilian:2007gr,Moretti:2001zz,Kilian:2011ka}\footnote{Note
  that it is not possible to use an automated tool for Feynman rules
  to include these rules (like e.g. via~\cite{Christensen:2010wz}) as
  one also needs a prescription to single out $s$ channels.}. This
allows us to numerically compute unitarized cross sections and
generate corresponding event samples at colliders.  

We note that up to the perturbative order that we are calculating, there is no
difference between the \textit{T}-matrix and \textit{K}-matrix unitarization prescriptions.
A difference would show up for higher-order or model-specific amplitudes that
initially contain an imaginary part.

\begin{figure}[h!]
  \centering
  \includegraphics[width=0.7\linewidth]{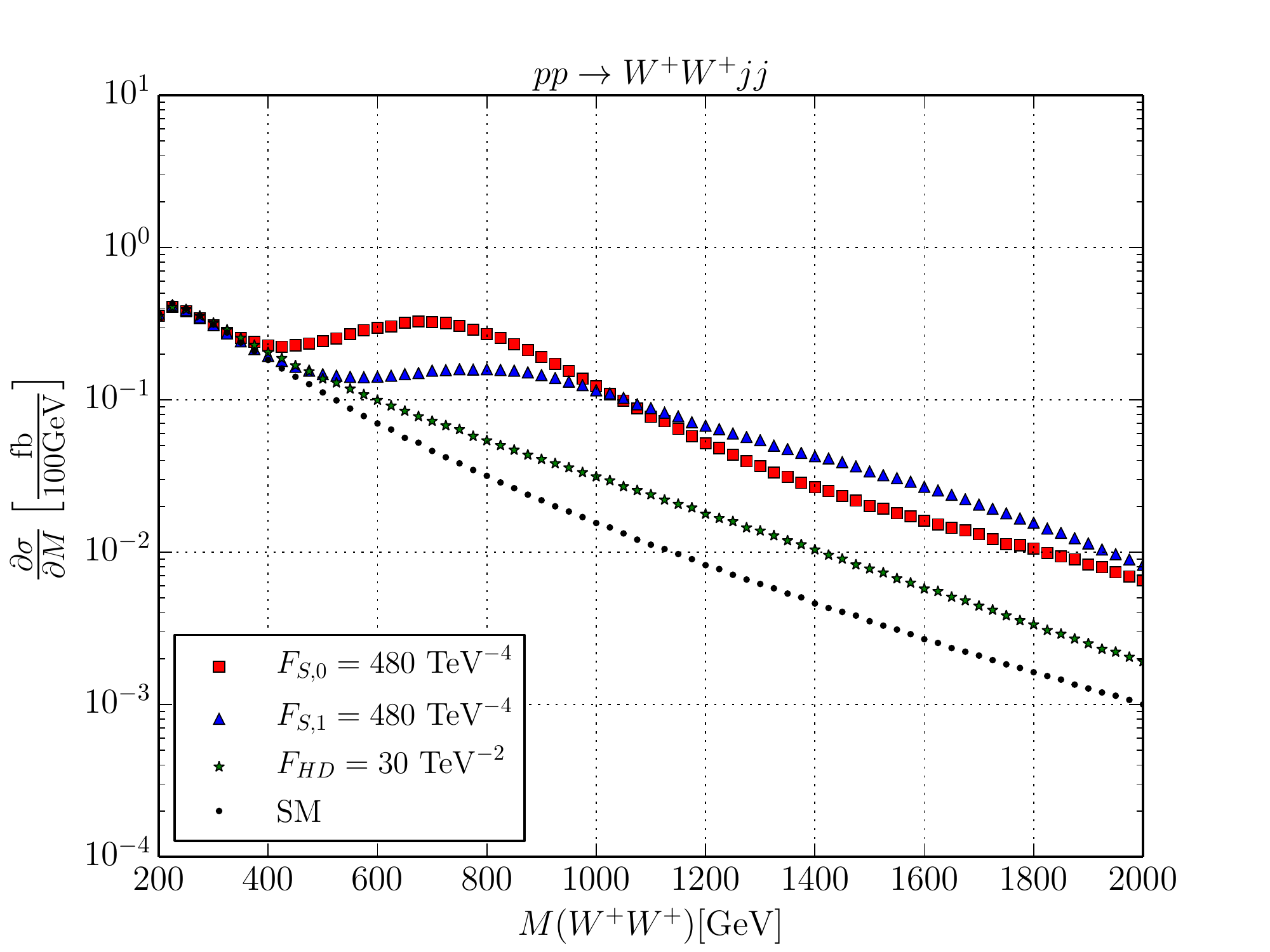}
  \caption{$pp\to W^+W^+jj$, unitarized
   (QCD contributions neglected).\\ Cuts: $M_{jj} > 500$ GeV;
    $\Delta\eta_{jj} > 2.4$;
    $p_T^j > 20$ GeV;
    $|\eta_j| < 4.5$.}
  \label{fig:W+W+}
\end{figure}

\begin{figure}[h!]
  \centering
  \includegraphics[width=0.7\linewidth]{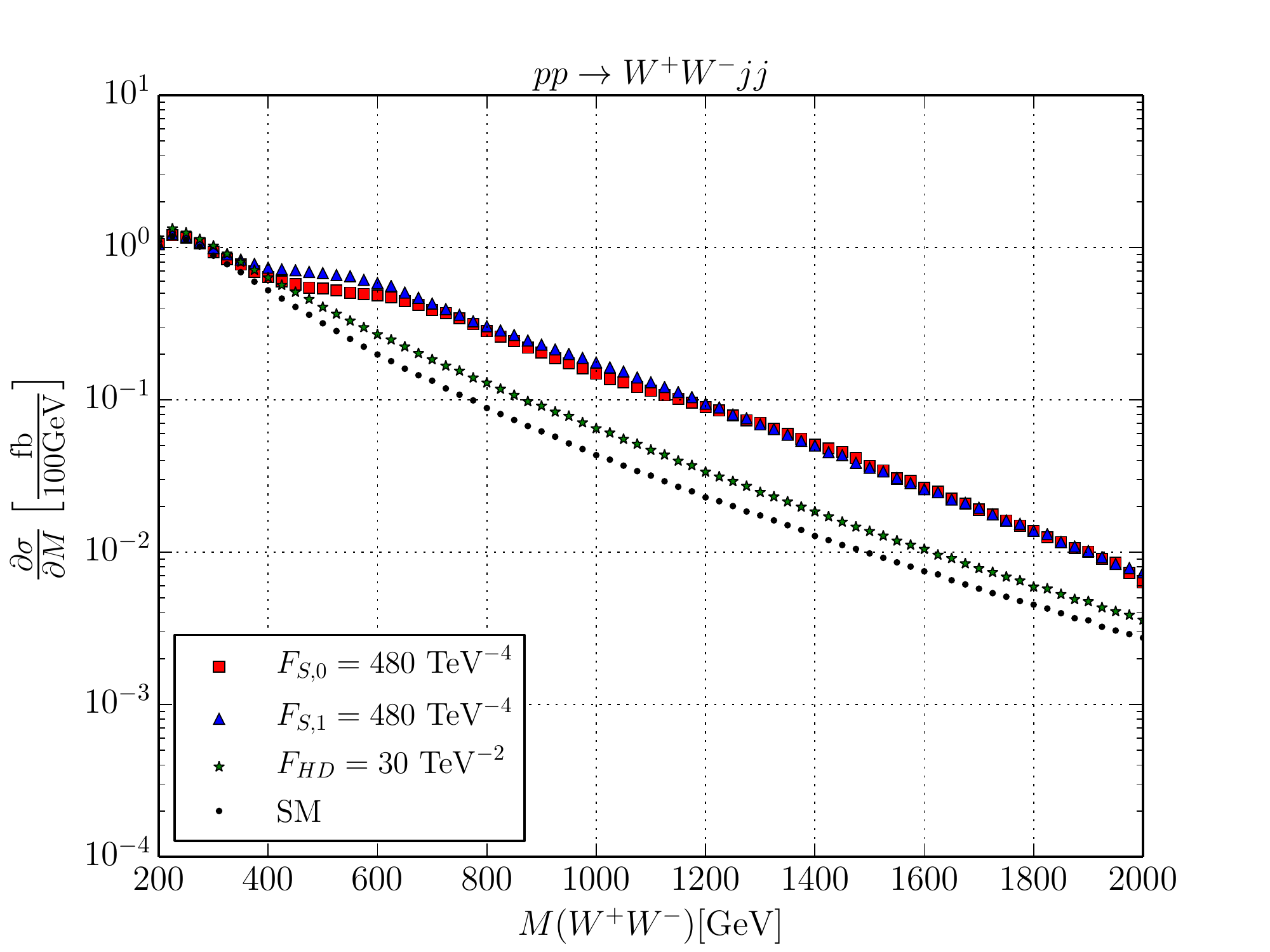}
  \caption{$pp\to W^+W^-jj$, unitarized
  (QCD contributions neglected).\\ Cuts: $M_{jj} > 500$ GeV;
    $\Delta\eta_{jj} > 2.4$;
    $p_T^j > 20$ GeV;
    $|\eta_j| < 4.5$.}
  \label{fig:W+W-}
\end{figure}

\begin{figure}[h!]
  \centering
  \includegraphics[width=0.7\linewidth]{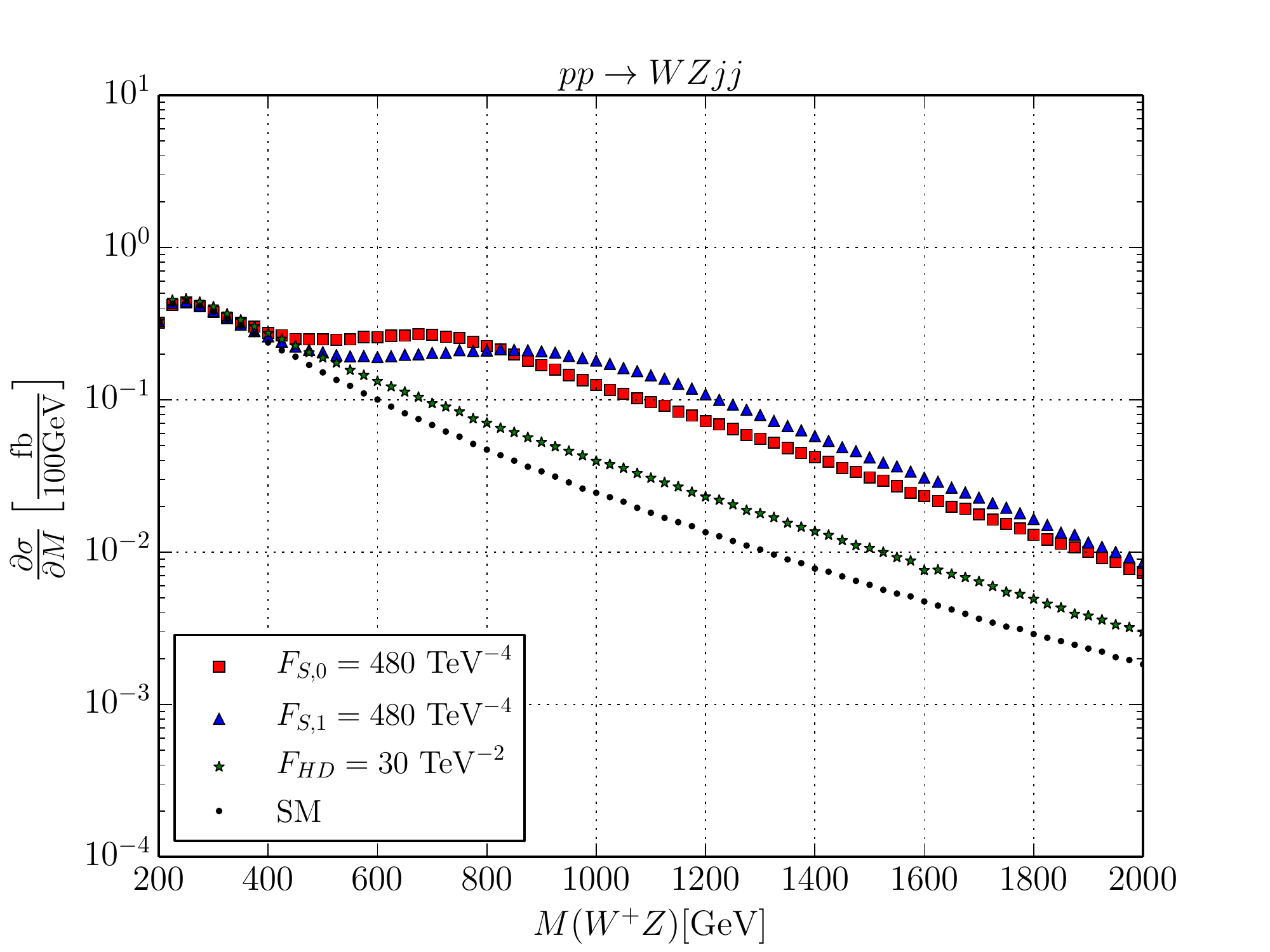}
  \caption{$pp\to W^+Zjj$, unitarized (QCD contributions neglected).
  \\ Cuts: $M_{jj} > 500$ GeV;
    $\Delta\eta_{jj} > 2.4$;
    $p_T^j > 20$ GeV;
    $|\eta_j| < 4.5$.}
  \label{fig:WZ}
\end{figure}

\begin{figure}[h!]
  \centering
  \includegraphics[width=0.7\linewidth]{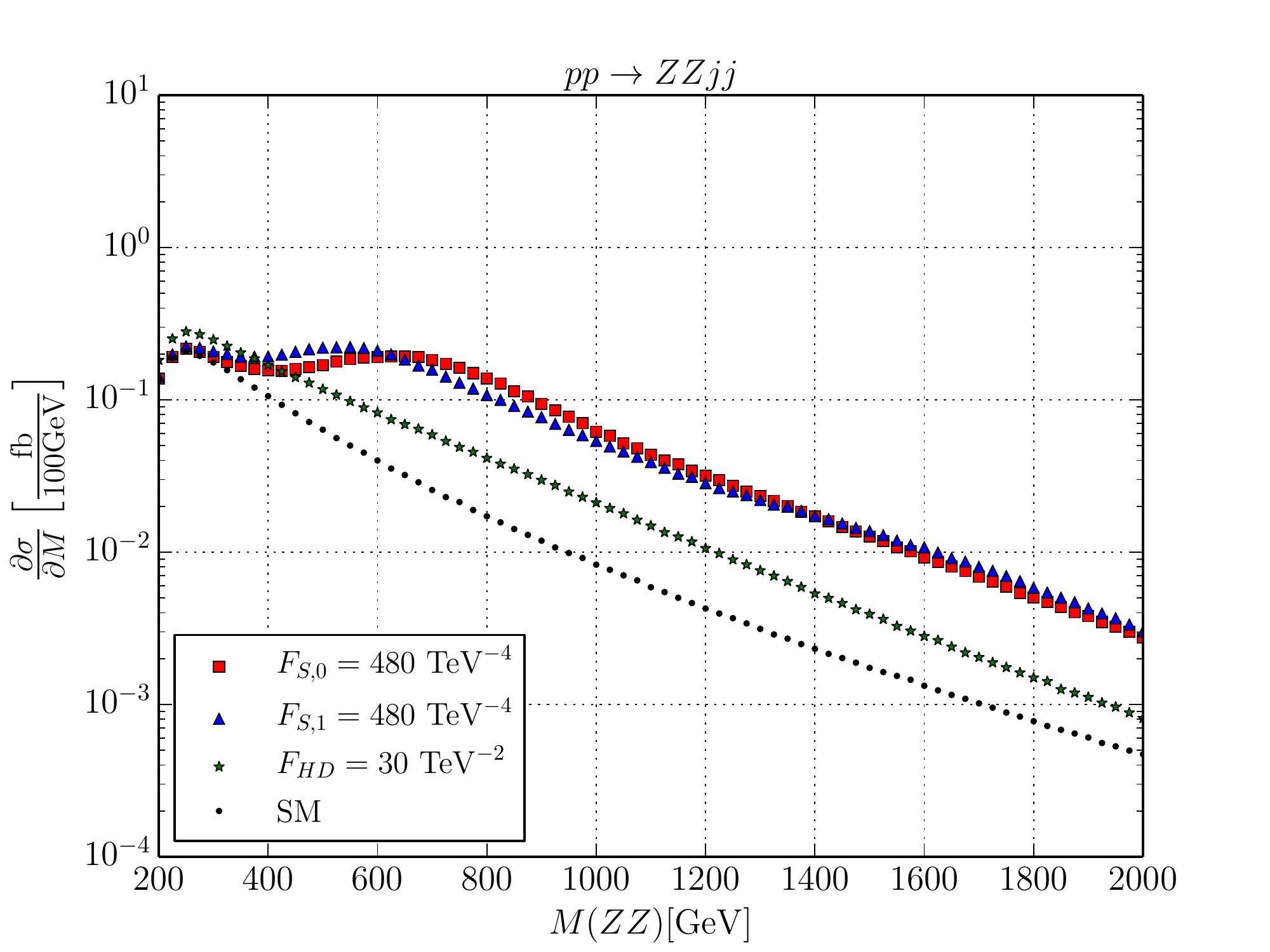}
  \caption{$pp\to ZZjj$, unitarized
  (QCD contributions neglected). \\ Cuts: $M_{jj} > 500$ GeV;
    $\Delta\eta_{jj} > 2.4$;
    $p_T^j > 20$ GeV;
    $|\eta_j| < 4.5$.}
  \label{fig:ZZ}
\end{figure}

The results in Figs.~\ref{fig:W+W+} to~\ref{fig:ZZ} are complementary
to Fig.~\ref{fig:W+W+bare}.  They display the
unitarized distribution of the VV invariant mass for the same selected
values of the parameters $F_{HD},F_{S,0},F_{S,1}$, again calculated
for the LHC configuration with $\sqrt{s}=14\;\TeV$ and standard cuts, 
dijet invariant mass $M_{jj} > 500$ GeV, jet rapidity distance
$\Delta\eta_{jj} > 2.4$, a minimal jet transverse momentum of $p_T >
20$ GeV, and a minimal (and opposite) jet rapidity of $|\eta_j| < 4.5$.
We show the distinct final states $W^+W^+$, $W^+W^-$, $W^+Z$, and $ZZ$
with the final-state vector bosons taken on-shell.

The plots clearly indicate that the naively calculated numbers with
anomalous couplings and no unitarization grossly overshoot the more
realistic \textit{T}-matrix results.  For the chosen parameters, the effect
of the dimension-eight operators is more pronounced than the effect of
the anomalous Higgs coupling, a dimension-six operator.  In all
channels, the unitarized curves fall down with energy with the same
rate as the SM curves, but enhanced by about one order of magnitude.
There is a distinct threshold region where the cross section
interpolates between the SM curve and the saturated limit.  Only
within this small window, a pure EFT description could be meaningful.

\subsection{Numerical Results: Full Processes}

At the LHC, the actual final state consists of six fermions, namely
\begin{figure}[H]
  \subfigure[$F_{HD}=30 \;
  \mathrm{TeV}^{-2}$]{\includegraphics[width=0.45\linewidth]{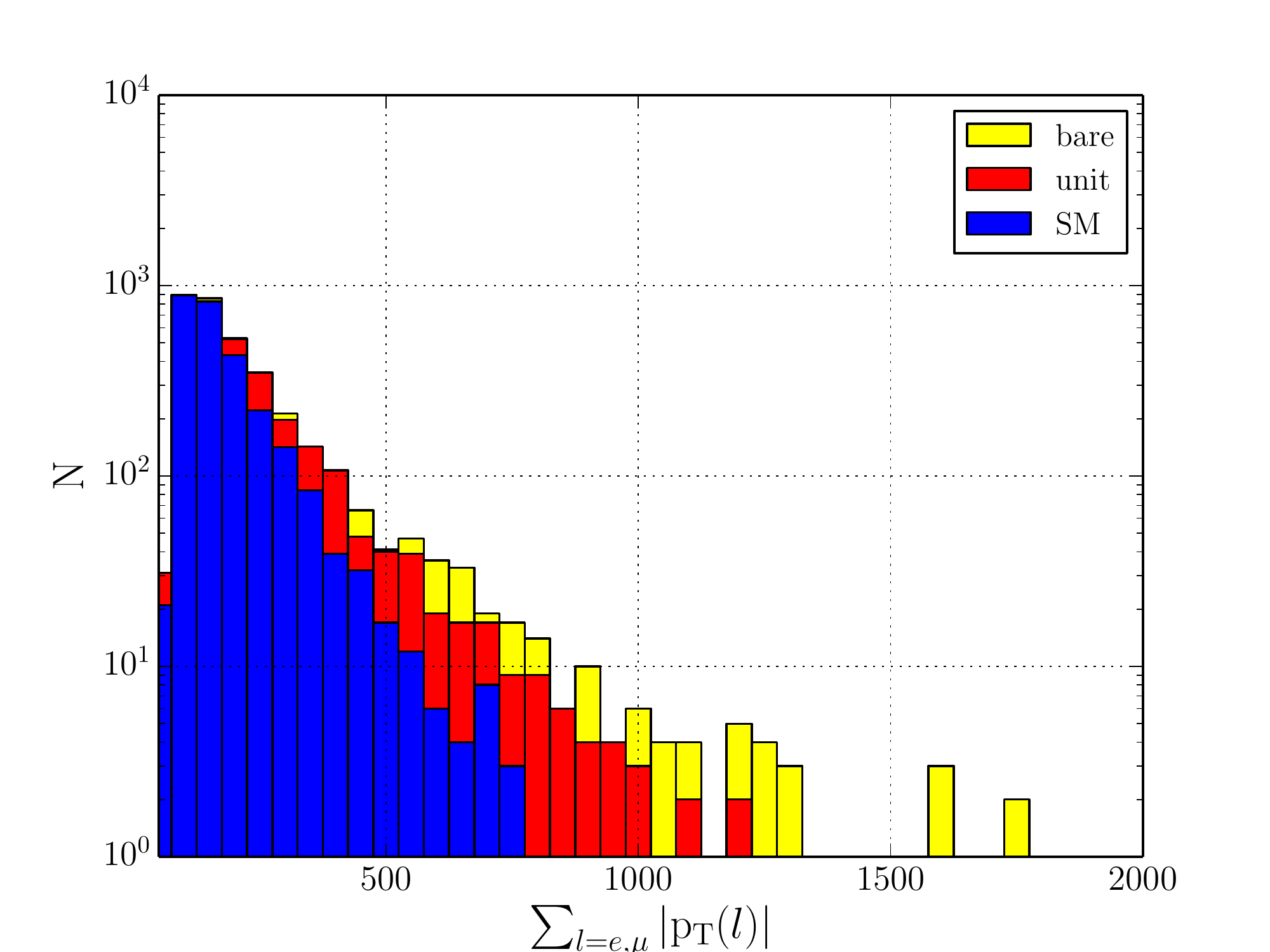} 
    \includegraphics[width=0.45\linewidth]{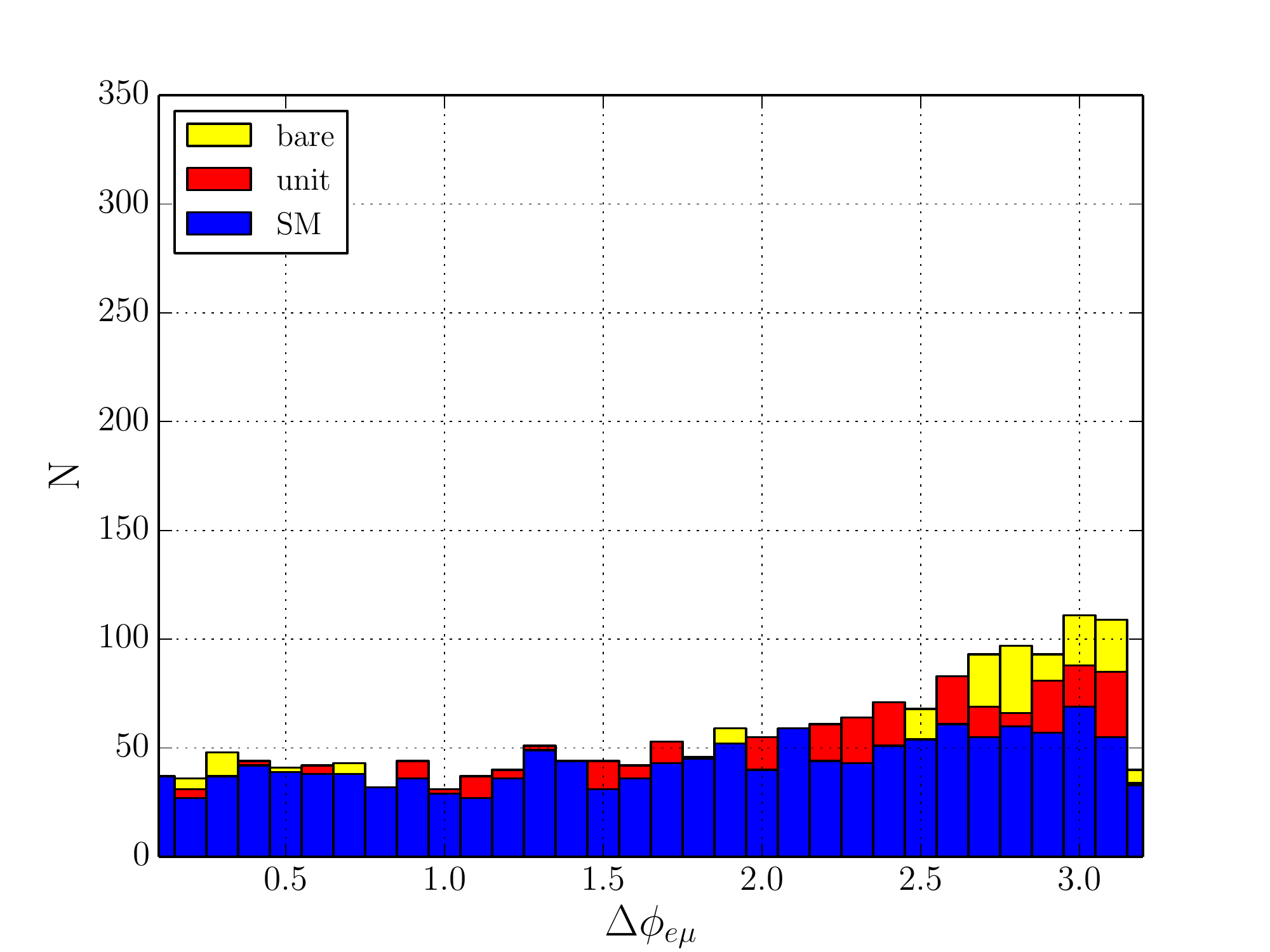}}\\
  \subfigure[$F_{S,0}=480 \; \mathrm{TeV}^{-4}$]{\includegraphics[width=0.45\linewidth]{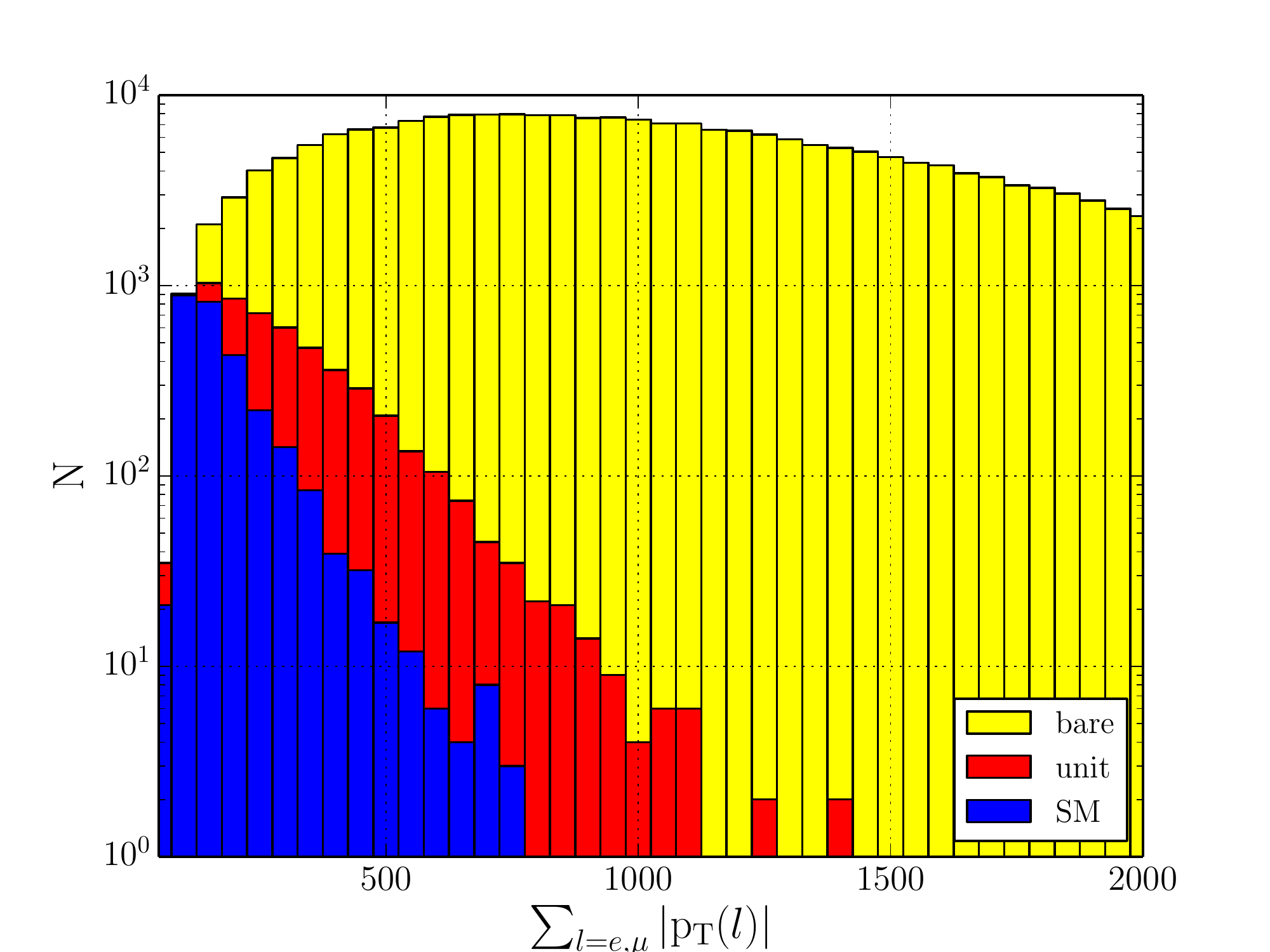}
    \includegraphics[width=0.45\linewidth]{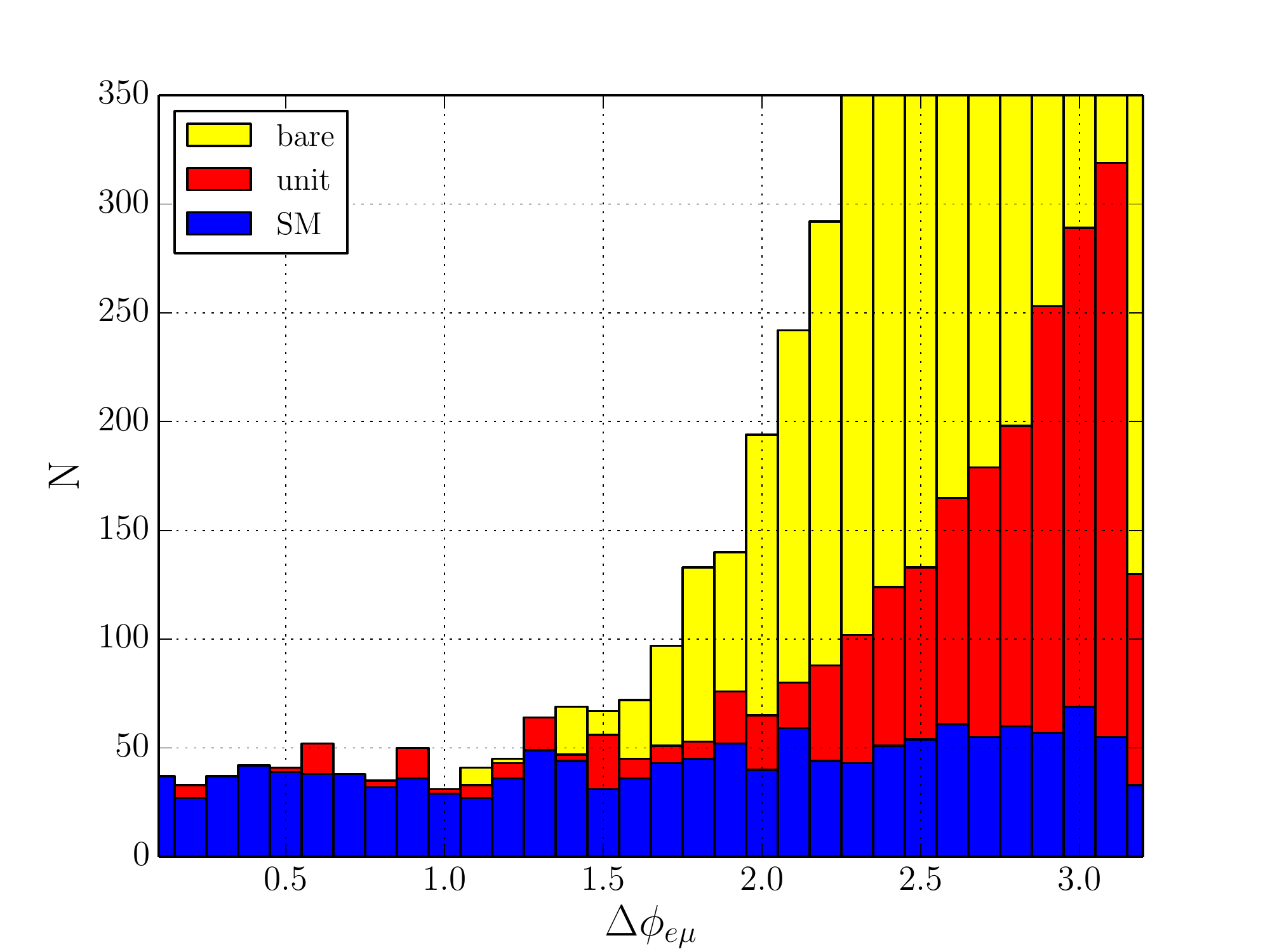}}\\
  \subfigure[$F_{S,1}=480 \; \mathrm{TeV}^{-4}$]{\includegraphics[width=0.45\linewidth]{a4-pt-log.pdf}
    \includegraphics[width=0.45\linewidth]{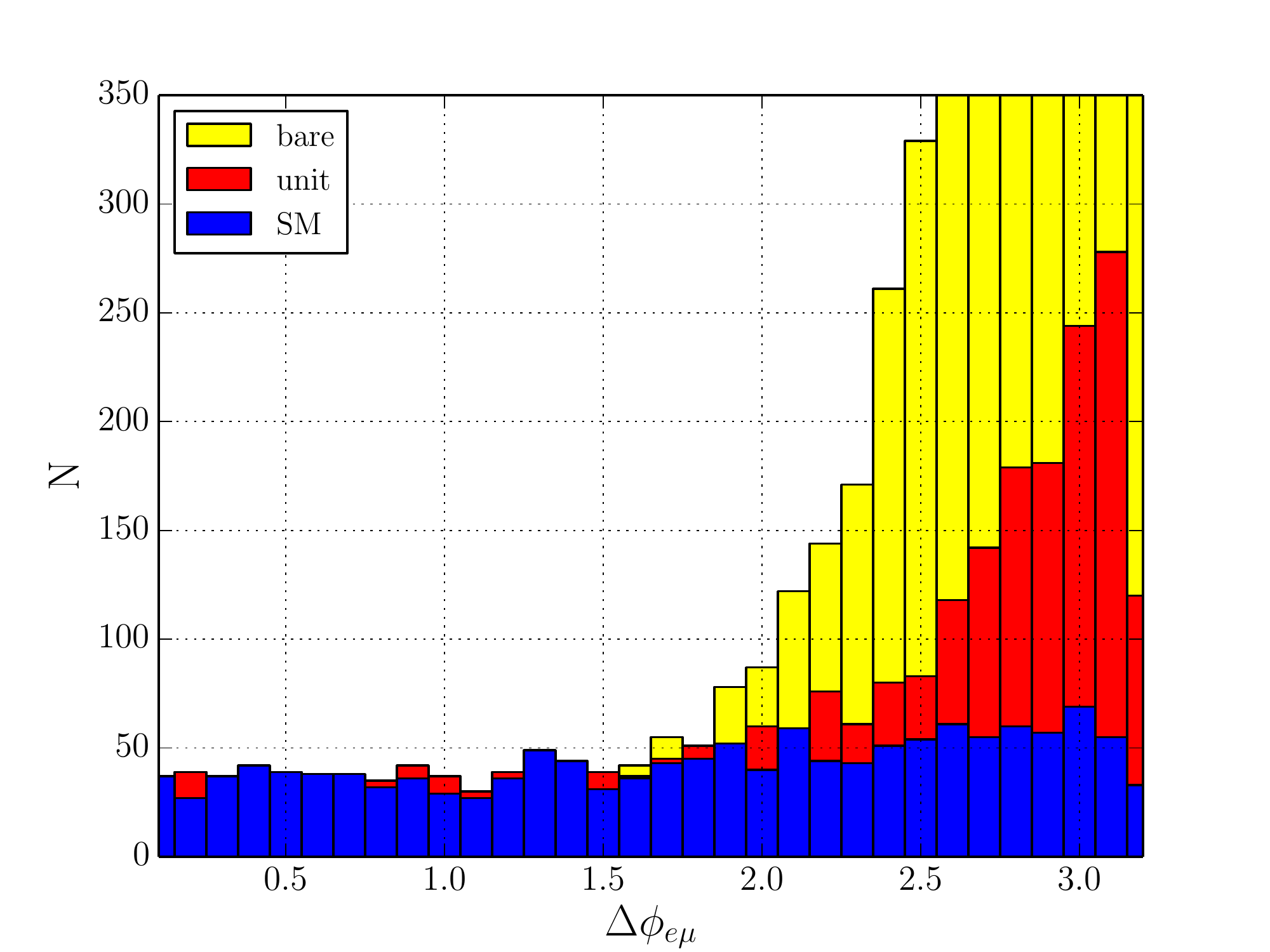}}\\
  \caption{$\mathrm{pp}\to e^+\mu^+\nu_e\nu_\mu jj, \sqrt{s}=14 \, \mathrm{TeV}, \mathrm{L}=1000 \, \mathrm{fb}^{-1}$  \\
    Cuts:  $M_{jj} > 500$ GeV;
    $\Delta\eta_{jj} > 2.4$;
    $p_T^j > 20 \, \mathrm{GeV}$;
    $|\eta_j| < 4.5$;
    $p_T^\ell > 20 \, \mathrm{GeV}$}
  \label{fig:6f-same-sign-pt-phi}
\end{figure}
two forward jets and the decay products of the vector bosons.  We
present results for the process with same-sign charged leptons,
\begin{equation}
  pp \to e^+\mu^+\nu_e\nu_\mu jj
\end{equation}
including the complete irreducible background.  The events have been
generated on the basis of the complete tree-level amplitude that
connects the initial and final state. The plots show an
unweighted partonic event sample that corresponds to
$1\;\mathrm{ab}^{-1}$ at the nominal LHC energy of $14\;\TeV$.  We
have applied standard VBF cuts, as listed in the figure captions.

In Fig.~\ref{fig:6f-same-sign-pt-phi}, we
show the scalar sum of transverse momentum and the azimuthal distance
of the charged lepton pair, respectively.  Both observables are
sensitive to the chosen values of the anomalous couplings.  There is a
significant difference between the SM prediction (blue/dark) and the
prediction with nonzero operator coefficient and unitarization
(red/medium). For reference, we also display the unphysical results
that we would generate without unitarization (yellow/light).

All numbers have been calculated with the \texttt{WHIZARD} event
generator~\cite{Kilian:2007gr} in version 2.2 which implements the
\textit{T}-matrix unitarized model with dimension-six and -eight operators,
as explained above.

\section{Summary and Conclusion}
\label{sec:conclusion}

We have developed a method to model the high-energy behavior of
quasi-elastic vector-boson scattering processes in a way that it can
be applied to collider analyses, covering in particular hadron
colliders where observables cannot always be limited to a narrow
energy range.  The method interpolates between the SM with a light Higgs
boson as the low-energy limit, its effective-theory extension,
and a high-energy behavior that remains consistent with unitarity
constraints.

It turns out that the only experimentally distinguishable
possibilities for vector-boson scattering process are (i) the pure SM,
(ii) new particles, as, e.g., in a two-Higgs doublet model, or (iii) a
deviation that smoothly increases with energy and indicates a strongly
interacting Higgs sector.  We study the latter possibility.  Small
deviations that stay within the weakly interacting regime are
mostly indistinguishable from the SM, at least in vector boson
scattering. 

In this work we do not propose any concrete model beyond the SM.
However, for quantitatively establishing the validity of the SM, or
for qualifying the significance of any possible experimental
discrepancy, we need an EFT approach that provides
parameterizations for deviations in all possible directions in model
space.  For being phenomenologically useful, such alternative
parameterizations must be consistent with unitarity as a limitation to
the number of events that can reasonably contribute to a particular
observable.

The problem of modelling high-energy electrowek interactions has
already been discussed three decades ago when multi-TeV colliders were
planned for the first time.  However, the present context is somewhat
different: a reasonable model must smoothly interpolate high-energy
strong interactions with the now-established \emph{light-}Higgs
scenario.  Adapting methods originally developed for the Higgs-less
case, we propose to unitarize the EFT amplitudes by extending the
parameter-free \textit{K}-matrix formalism.  We reformulate this method as a direct
\textit{T}-matrix scheme, such
that it unitarizes any given model without requiring a perturbative
expansion or introducing additional structure in the result.  We have
described this approach in detail, including the systematic embedding
of the new effects in the machinery of Monte-Carlo simulation for the
full multi-fermion processes.

The underlying \textit{T}-matrix prescription ensures that any computed results
do not overshoot the physical limit, but it does not have any further
physical interpretation.  Given sufficient experimental precision, we
should get a handle on the behavior of the invariant-mass distribution
beyond the maximum that is related to the strong-interaction
threshold.  Possibilities for modelling VBS beyond
this threshold have been sketched
in~\cite{Reuter:2013gla,Reuter:2014kya} and will be developed in more
detail in a separate paper~\cite{Kilian:2015opv}.


\subsection*{Acknowledgments}
We acknowledge enlightening discussions and helpful
input from Philipp Anger, Carsten Bittrich, Tao Han, Michael Kobel,
Ashutosh Kotwal, Marc-Andr\'{e} Pleier, Ulrike Schnoor, and Anja
Vest. JRR has been partially supported by the Strategic Helmholtz
Alliance 'Physics at the Terascale' of the German HGF.


\appendix
\section{Notational Conventions}
\label{sec:notation}

The field content of the EFT is given by fermions, gluons, electroweak
vector bosons, and the Higgs doublet which in a linear gauge consists
of the physical Higgs boson and three Goldstone bosons $w^+,w^-,w^3$.
We do not write fermions or gluons explicitly.  For electroweak gauge
bosons, we define
\begin{align}
  \label{covar-D}
  \vD_\mu \vH 
  &= \partial_\mu \vH - \ii g \vW_\mu \vH - \ii g^\prime \vH \vB_\mu 
  \\
  \label{W-field}
  \vW_{\mu\nu}
  &=\partial_\mu \vW_\nu - \partial_\nu \vW_\mu 
  - \ii g \left [ \vW_\mu , \vW_\nu \right ] 
  \\
  \vB_{\mu\nu}
  &=\partial_\mu \vB_\nu - \partial_\nu \vB_\mu
\end{align}
with
\begin{align}
  \vW_{\mu} &= W_\mu^a \frac{\tau^a}{2}, &
  \vB_{\mu} &= -\frac{\tau^3}{2} B_\mu 
\end{align}
In the linear representation, the SM Higgs field combines with the
Goldstone bosons as an electroweak doublet.  The Higgs sector has an
additional global $SU(2)_C$ (custodial) symmetry~\cite{Sikivie:1980hm}.  In
order to make the $SU(2)_C$ transformation properties explicit, we
parameterize the Higgs field as the Hermitian matrix
\begin{equation}
  \vH =
  \frac 1 2 
  \begin{pmatrix}
    v + h -\ii w^3 & -\ii \sqrt{2} w^+ \\
    -\ii \sqrt{2} w^- & v + h + \ii w^3  \\
  \end{pmatrix} \qquad .
\end{equation}
The physical Higgs field multiplies the unit matrix, while the Goldstone
bosons $w^+,w^3,w^-$ are the components proportional to the Pauli matrices
$\tau^+,\tau^3,\tau^-$.  $SU(2)_L$ transformations $U_L$ and $SU(2)_R$
transformations $U_R$, and custodial $SU(2)_C$ transformations $U_C$ act as
\begin{align}
  \vH &\to U_L\vH, &
  \vH &\to \vH U_R^\dagger, &
  \vH &\to U_C\vH U_C^\dagger,
\end{align}
respectively.  The $\tau^3$ part of $SU(2)_R$ coincides with
hypercharge $U(1)_Y$ transformations, while $\tau^{1,2}$-associated
transformations are not realized as local gauge symmetries.  Under
custodial trunsformations, the Higgs field decomposes into singlet
(the physical Higgs) and triplet (Goldstones).  Conversely, under
$SU(2)_L$ gauge transformations, the two columns of the Higgs matrix
transform independently as the conventional complex doublet $\Phi$ and
its charge conjugate.  In unitarity gauge, the Goldstone bosons
disappear, and the matrix reduces to the $v+h$ term.

The bosonic part of the lowest order EFT, i.e., the plain SM
Lagrangian, reads
\begin{align}
\label{SM-LO}
  \LL_{\text{min}}=&-\frac{1}{2}\tr{\vW_{\mu\nu}\vW^{\mu\nu}}
		-\frac{1}{2}\tr{\vB_{\mu\nu}\vB^{\mu\nu}} \\
		&+\tr{ \left ( \vD_\mu \vH \right )^\dagger
		 \vD^\mu \vH }
		+ \mu^2\tr{\vH^\dagger \vH}
		-\frac{\lambda}{2}\left( \tr{\vH^\dagger \vH} \right)^2,
\end{align}
For a precise definition of higher-dimensional operators, we have to
express the free parameters of the EFT, order by order in the operator
dimension, in terms of observable quantities.  A possible choice for
such a renormalization scheme, applicable to the operator expansion at
tree-level and beyond, is
\begin{align}
  g &= 2\frac{m_W}{v}, &
  g' &= 2\frac{\sqrt{m_Z^2-m_W^2}}{v}, &
  \mu^2 &= \frac12 m_H^2, &
  \lambda &= \frac{m_H^2}{v^2}.
\end{align}
for the parameters in the SM Lagrangian, (\ref{SM-LO}), with particle
masses and the Higgs vacuum expectation value $v$ as fixed input.  In
particular, the definition of $g$ and $g'$ unambiguously determines
the covariant field strength and the covariant derivative that we use
for constructing higher-dimensional operators.  Furthermore, we may
fix the kinetic-energy normalization to their conventional SM values.

We have deliberately excluded fermions here.  Light fermions are
coupled by gauge bosons.  For our purposes, they act like external
currents, and are properly taken into account when the unitarized
amplitudes are embedded into the full process.  Heavy fermions are
important in the context of Higgs physics, but absent from the initial
state.  In the final state, they are identifiable.  Here, we just
consider processes which ultimately involve light fermions.  In
passing, we note that genuine anomalous interactions of light fermions
are experimentally accessible in processes such as lepton and jet pair
production.


\section{Unitarization, K-Matrix, and All That: Proofs}
\label{sec:kmatrixunit}

\subsection{Non-Hermitian \textit{K}-Matrix}

If the \textit{K}-matrix is not Hermitian, we need to find a generalization
of~\eqref{eq:Cayley}, i.\,e.
\begin{equation}
  T = \frac{K'}{\mathbf{1} - \ii K'/2}
\end{equation}
with a suitable~$K'$. The most straightforward approach is to 
just throw away the imaginary parts~$K'=\Re K=(K+K^\dagger)/2$.  The
interpretation of the Cayley transform as an inverse stereographic
projection suggests a less drastic approach, that retains  the
imaginary part.  Consider the family~$\{K_\kappa\}$ of \textit{K}-matrices
that have the same projection with center~$\ii\mathbf{1}$
\begin{subequations}
\begin{align}
\label{eq:K'-scaling}
  \frac{K_\kappa}{2} - \ii\mathbf{1}
     &= \kappa \left(\frac{K}{2}-\ii\mathbf{1}\right) \kappa \\
\label{eq:kappa-hermitian}
  \kappa^\dagger &= \kappa > 0 \,.
\end{align}
and choose the unique self adjoint member~$K'\in\{K_\kappa\}$ of this
family 
\begin{equation}
\label{eq:K'-hermitian}
  K' = (K')^\dagger\,.
\end{equation}
\end{subequations}
As long\footnote{%
  We can use the Riesz-Dunford functional calculus~ \cite{Riesz:1930,Gelfand:1941,Dunford:1988}
  to construct projectors on subspaces corresponding to parts
  of the spectrum of~$\Im K/2$
  \begin{equation}
    P_\Sigma = \int_{\partial\Sigma}\frac{\dd z}{2\pi\ii}
    \frac{1}{z\mathbf{1} - \Im K/2}\,,
  \end{equation}
  where~$\Sigma$ contains the desired part of the spectrum of~$\Im K/2$.}
as $\Im K/2 < \mathbf{1}$, there is a unique solution with a
converging power series expansion
\begin{subequations}
\begin{align}
\label{eq:kappa}
  \kappa &= \frac{1}{\sqrt{\mathbf{1} - \Im K/2}} \\
\label{eq:K'}
  K' &= \kappa (\Re K) \kappa
\end{align}
\end{subequations}
resulting in
\begin{equation}
  T = \kappa (\Re K) \frac{1}{\mathbf{1} - \ii K^\dagger/2} \kappa^{-1}
    = \kappa^{-1} \frac{1}{\mathbf{1} - \ii K^\dagger/2} (\Re K)  \kappa
\end{equation}
For normal~\textit{K}, i.\,e.~$KK^\dagger=K^\dagger K$ everything commutes
and we may write 
\begin{equation*}
  T = \frac{\Re K}{\mathbf{1}-\ii K^\dagger/2}
    = \frac{\Re K}{\mathbf{1}-\ii \Re K/2 - \Im K/2}
\end{equation*}
instead, highlighting the contribution of~$\Im K=(K-K^\dagger)/2\ii$.


\subsection{Properties of \textit{T}-matrix unitarized (Linear Projection) operators}

The unitarity of the $S$ matrix, $SS^\dagger = S^\dagger S = 1$
implies that each interaction matrix, $S = 1 + i T$, has to satisfy 
\begin{align}
	\text{T}^\dagger \text{T} =
		 -\ii \left (\text{T} -\text{T}^\dagger \right ).
\end{align}

For \textit{T}-matrix unitarized operators~\eqref{eq:T(T0)} via linear projection, we use
\begin{equation}
  \mathrm{T}\left ( \mathrm{T}_0 \right ) 
  = \frac{\mathrm{Re} \mathrm{T}_0}
    {\mathds{1} -\frac{\ii}{2}\mathrm{T}_0^\dagger } \\ 
  = \frac{\mathrm{Re} \mathrm{T}_0}{\mathds{1} 
    +\frac{1}{4}\mathrm{T}_0\mathrm{T}_0^\dagger} 
   \left( \mathds{1} + \frac{\ii}{2} \mathrm{T}_0 \right)
\end{equation}
to show the unitarity of the corresponding $S$ operator:
\begin{equation}
  \begin{aligned}
    \mathrm{SS}^\dagger&= \mathds{1}-2 \mathrm{Im} \left (T \right)+\mathrm{TT}^\dagger \\
    &=\mathds{1}-\frac{\left(\mathrm{Re}
        \mathrm{T}_0\right)^2}{\mathds{1} 
    +\frac{1}{4}\mathrm{T}_0\mathrm{T}_0^\dagger} + 
    \frac{\left(\mathrm{Re}
        \mathrm{T}_0\right)^2}{\mathds{1} 
    +\frac{1}{4}\mathrm{T}_0\mathrm{T}_0^\dagger}
   = \mathds{1}.
  \end{aligned}
\end{equation}

In the same way, we can show the idempotency of the \textit{T} operation:
\begin{equation}
  \begin{aligned}
    \mathrm{T}\left ( \mathrm{T} \left ( \mathrm{T}_0 \right )  \right )
    &= \frac{\mathrm{Re} \mathrm{T}(\mathrm{T}_0)}{1 - \frac{\ii}{2}
      \mathrm{T}(\mathrm{T}_0)^\dagger} = 
    \frac{\frac{\mathrm{Re}
        \mathrm{T}_0}{\mathds{1} 
    +\frac{1}{4}\mathrm{T}_0\mathrm{T}_0^\dagger} \left( \mathds{1} -
    \frac{1}{2} \mathrm{Im} \mathrm{T}_0 \right)}{\mathds{1} -
  \frac{\ii}{2} \frac{\mathrm{Re} \mathrm{T}_0 \; \left( \mathds{1} -
      \frac{\ii}{2} 
      \mathrm{T}_0^\dagger \right)}{\left( \mathds{1} - \frac{\ii}{2}
      \mathrm{T}_0^\dagger \right)\left( \mathds{1} + \frac{\ii}{2}
      \mathrm{T}_0 \right)}} \\ 
    &= \frac{\mathrm{Re}\mathrm{T}_0}{\mathds{1} 
    +\frac{1}{4}\mathrm{T}_0\mathrm{T}_0^\dagger} \left( \mathds{1} +
    \frac{\ii}{2} \mathrm{T}_0 \right) = \mathrm{T} (\mathrm{T}_0)
  \end{aligned}
\end{equation}


\subsection{Properties of \textit{T}-matrix unitarized (Thales Projection) operators }

In this section we take the definition of the \textit{T}-matrix unitarized
operator from~\eqref{eq:generalop}, and show, using

\begin{equation}
  \mathrm{T}\left ( \mathrm{T}_0 \right ) 
  = \frac{1}{\mathrm{Re} \left (\frac{1}{\mathrm{T}_0}\right
    )-\frac{\ii}{2}\mathds{1} } \\ 
  = \frac{1}{\mathrm{Re} \left (\frac{1}{\mathrm{T}_0}\right
    )^2+\frac{1}{4}\mathds{1}} 
  \left (\mathrm{Re} \left(\frac{1}{\mathrm{T}_0} \right
    )+\frac{\ii}{2}\mathds{1} \right )  \qquad ,
\end{equation}
the unitarity of the corresponding $S$ operator:
\begin{equation}
  \begin{aligned}
    \mathrm{SS}^\dagger&= \mathds{1}-2 \mathrm{Im} \left (T \right)+\mathrm{TT}^\dagger \\
    &=\mathds{1}-\frac{1}{\mathrm{Re} \left (\frac{1}{\mathrm{T}_0}\right )^2+\frac{1}{4}\mathds{1}}
    +\frac{1}{\mathrm{Re} \left (\frac{1}{\mathrm{T}_0}\right )^2+\frac{1}{4}\mathds{1}}
    = \mathds{1}.
  \end{aligned}
\end{equation}

Also, it is easy to see that this operation is idempotent:
\begin{equation}
  \begin{aligned}
    \mathrm{T}\left ( \mathrm{T} \left ( \mathrm{T}_0 \right )  \right )
    &= \frac{1}{\mathrm{Re} \left (\frac{1}{\mathrm{T} \left ( \mathrm{T}_0 \right )}\right )-\frac{\ii}{2}\mathds{1} }
    = \frac{1}{\mathrm{Re} \left( 
        \mathrm{Re} \left (\frac{1}{\mathrm{T}_0}\right ) - \frac{\ii}{2} \mathds{1} \right) -\frac{\ii}{2}\mathds{1}
    } \\
    &= \frac{1}{\mathrm{Re} \left (\frac{1}{\mathrm{T}_0}\right )-\frac{\ii}{2}\mathds{1} }
    = \mathrm{T}\left ( \mathrm{T}_0 \right ) .
  \end{aligned}
\end{equation}


\section{Operator Bases and their Translation}

\subsection{Introduction to different sets of operator bases}

It has become customary to write the EFT operator basis in a form that
is algebraically simple, so each basic operator is a single monomial
of the fields with a single coefficient.  For the renormalizable part
of the theory, this is justified by the usual renormalization
procedure where all terms are renormalized multiplicatively.

There is a vast literature on choices of operator bases for
dimension-6 and -8 operators in the electroweak sector; we only need a
sample operator here to demonstrate our point about the unitarization
procedure, so we only briefly mention the translation between the
non-linear and linear matrix representation of these operators. An
extensive discussion of the operator bases is a different topic and
discussed in a follow-up paper~\cite{Kilian:2015opv}.


\subsection{Translation between Nonlinear and Linear Matrix
  representation} 
We can compare the effective Lagrangians in Appelquist-Alboteanu
parameterization 
\begin{align*}
  \LL_4 &= \alpha_4 \Tr \left [ \vV_\mu \vV_\nu\right ]\Tr 
  \left [ \vV^\mu \vV^\nu\right ], \\
  \LL_5 &= \alpha_5 \Tr \left [ \vV_\mu \vV^\mu\right ]\Tr 
  \left [ \vV_\nu \vV^\nu\right ]
\end{align*}
from \cite{Appelquist:1980vg,Alboteanu:2008my} in unitarity gauge 
\begin{align*}
  \mathbf{V}_{\mu}&\hat{=} -\ii gW_\mu^a\frac{\tau^a}{2}
  + \ii g^\prime B_\mu\frac{\tau^3}{2}\\
\end{align*}  
to $\LL_{S,0}$ and $\LL_{S,1}$. Because we are only interested in 
the VBS part of these two Lagrangian, simplifying the covariant
derivative from $\LL_{S,0}$ and $\LL_{S,1}$ as
\begin{align}
  \vD_\mu \vH 
  &=  \frac{v}{2} \left (- \ii g \vW_\mu - \ii g^\prime \vB_\mu \right
  ) \notag = \frac{v}{2} \left ( -\ii gW_\mu^a\frac{\tau^a}{2}
    + \ii g^\prime B_\mu\frac{\tau^3}{2} \right ) \notag \\
  &= \frac{v}{2}\mathbf{V}_{\mu}, \\
  \left (\vD_\mu \vH  \right )^\dagger &= - \frac{v}{2}\mathbf{V}_{\mu}.
\end{align}
leads to
\begin{alignat}{3}
  \label{LL-S0x}
	\LL_{S,0}&=
	 & &F_{S,0}\frac{v^4}{16}\ &&
	  \Tr \left [ \vV_\mu \vV_\nu\right ]\Tr \left [ \vV^\mu \vV^\nu\right ],
\\
  \label{LL-S1x}
	\LL_{S,1}&=
	 & &F_{S,1} \frac{v^4}{16}\ &&
	  \Tr \left [ \vV_\mu \vV^\mu\right ]\Tr \left [ \vV_\nu \vV^\nu\right ].
\end{alignat}
Therefore we can relate the coefficients of these different notations:
\begin{align}
	\alpha_4 &= F_{S,0}\frac{v^4}{16}, \\
	\alpha_5 &= F_{S,1}\frac{v^4}{16}.
\end{align}
So the coefficients for the operators $\LL_{S,0}$ and $\LL_{S,1}$ are
equivalent to values of $\alpha_4$ and $\alpha_5$ of $\sim 0.11$,
which are within the limits from the latest ATLAS analysis
\cite{Aad:2014zda} ($-0.14 < \alpha_4 < 0.16$ and $-0.23 < \alpha_5 <
0.24$).

\section{Feynman Rules}

\subsection{Feynman Rules from New Physics Operators}
\label{sec:Feynman-rules-new-Operators}

\subsubsection{$\LL_S$}

The Lagrangian

\begin{alignat}{3}
  \LL_{S,0}&=
  & &F_{S,0}\ &&
  \tr{ \left ( \vD_\mu \vH \right )^\dagger \vD_\nu \vH}
  \cdot \tr{ \left ( \vD^\mu \vH \right )^\dagger \vD^\nu \vH}
  \\
  \LL_{S,1}&=
  & &F_{S,1}\ &&
  \tr{ \left ( \vD_\mu \vH \right )^\dagger \vD^\mu \vH}
  \cdot \tr{ \left ( \vD_\nu \vH \right )^\dagger \vD^\nu \vH} 
\end{alignat}

leads to the following Feynman rules in the unitarity gauge
(neglecting all vertices including a Higgs boson and five or more
external fields):

\begin{alignat}{2}
  W^+_{\mu_1}W^+_{\mu_2}W^-_{\mu_3}W^-_{\mu_4} :& \hspace{0.5 cm} &
  \frac{ \ii g^4 v^4}{16}& \left [
    \left (F_{S,0} + 2F_{S,1} \right )  \left (
      g_{\mu_1\mu_3}g_{\mu_2\mu_4}+g_{\mu_1\mu_4}g_{\mu_2\mu_3}
    \right ) \right . \notag \\
  &&& \qquad \left . + 2 F_{S,0}  g_{\mu_1\mu_2}g_{\mu_3\mu_4}
  \right ]
  \\
  Z_{\mu_1}Z_{\mu_2}W^+_{\mu_3}W^-_{\mu_4} :& \hspace{0.5 cm} &
  \frac{ \ii g^4 v^4}{16 c_w^2} & \left [ F_{S,0} \left (
      g_{\mu_1\mu_3}g_{\mu_2\mu_4}+g_{\mu_1\mu_4}g_{\mu_2\mu_3}
    \right ) \right . \notag \\
  && & \qquad \left. +2 F_{S,1}  g_{\mu_1\mu_2}g_{\mu_3\mu_4} \right ]
  \\
  Z_{\mu_1}Z_{\mu_2}Z_{\mu_3}Z_{\mu_4} :& \hspace{0.5 cm} &
  \frac{ \ii g^4 v^4}{8 c_w^4}  \left ( F_{S,0} +F_{S,1} \right ) & \left (
    g_{\mu_1\mu_2}g_{\mu_3\mu_4} 
    + g_{\mu_1\mu_3}g_{\mu_2\mu_4}
    +g_{\mu_1\mu_4}g_{\mu_2\mu_3}
  \right )
\end{alignat}

\subsubsection{$\LL_{HD}$}
The Lagrangian
\begin{align}
	\LL_{HD}&=F_{HD} \tr{\vH^\dagger \vH - \frac{v^2}{4}} \cdot
		\tr{ \left ( \vD_\mu \vH \right )^\dagger \vD_\mu \vH}
\end{align}
leads to the following Feynman rules in unitarity gauge 
(neglecting all vertices with more than one Higgs):
\begin{alignat}{2}
  hW^+_\mu W^-_\nu:& \hspace{2 cm} &
  \frac{\ii g^2 v^3}{4} F_{HD} & g_{\mu\nu} \\
  hZ_\mu Z_\nu:& \hspace{2 cm} &
  \frac{\ii g^2 v^3}{4 s_w^2} F_{HD} & g_{\mu\nu} 
\end{alignat}

\subsection{Feynman Rules: Unitarization Corrections}
\label{sec:Feynman-rules}

These ``Feynman Rules'' are only used for s-channel scattering of
$VV \to VV$ with the center-of-mass energy $s= \left (p_1 + p_2 \right )^2$ and counterterms $\mathcal{A}_{ij}$(\ref{eq:amplitudeCT}) 
\begin{alignat}{3}
  W^{\pm}_{\mu_1}W^{\pm}_{\mu_2}\rightarrow W^{\pm}_{\mu_3}W^{\pm}_{\mu_4}
  &: \quad & & \frac{g^4 v^4}{4} & &\left [ 
  \left ( \Delta \amp_{02}(s)-10\Delta \amp_{22}(s) \right )
  \frac{g_{\mu_1\mu_2}g_{\mu_3\mu_4}}{s^2}
   \right .
  \notag\\
  &&&& & \left .+ 15 \Delta \amp_{22}(s)
  \frac{g_{\mu_1\mu_3}g_{\mu_2\mu_4}+ g_{\mu_1\mu_4}g_{\mu_2\mu_3}}{s^2} \right ]
  \\
  W^\pm_{\mu_1}W^\mp_{\mu_2}\rightarrow Z_{\mu_3}Z_{\mu_4}
  &:& & \frac{g^4 v^4}{4 c_w^2} & &\left [ 
  \left ( \frac{1}{3} \left ( \Delta \amp_{00}(s)-\Delta \amp_{20}(s)
  \right ) \right . \right . \notag \\
  &&&& &\left. \left . -\frac{10}{3}\left ( \Delta \amp_{02}(s)-\Delta \amp_{22}(s)
  \right )   \right )
  \frac{g_{\mu_1\mu_2}g_{\mu_3\mu_4}}{s^2}
   \right .
  \notag\\
  &&&& &\left . + 5 \left (\Delta \amp_{02}(s)-\Delta \amp_{22}(s) \right)
  \frac{g_{\mu_1\mu_3}g_{\mu_2\mu_4}+ g_{\mu_1\mu_4}g_{\mu_2\mu_3}}{s^2} \right ]
  \\
  W^\pm_{\mu_1}Z_{\mu_2}\rightarrow W^\pm_{\mu_3}Z_{\mu_4}
  &:& & \frac{g^4 v^4}{4c_w^2} & &\left [
  \left (\frac{1}{2}\Delta \amp_{20}(s)-5\Delta \amp_{22}(s) \right )
  \frac{g_{\mu_1\mu_2}g_{\mu_3\mu_4}}{s^2}
   \right .
  \notag\\
  &&&& &+ \left ( -\frac{3}{2}\Delta \amp_{11}(s)+\frac{15}{2}\Delta
    \amp_{22}(s) \right )\frac{g_{\mu_1\mu_3}g_{\mu_2\mu_4}}{s^2}
  \notag\\
  &&&& & \left .+  \left (\frac{3}{2}\Delta \amp_{11}(s)+\frac{15}{2}\Delta
    \amp_{22}(s) \right )\frac{g_{\mu_1\mu_4}g_{\mu_2\mu_3}}{s^2}
    \right ] 
\end{alignat}
\begin{alignat}{3}
  W^\pm_{\mu_1}W^\mp_{\mu_2}\rightarrow W^\pm_{\mu_3}W^\mp_{\mu_4}
  &:& & \frac{g^4 v^4}{4} & &\left [
  \left ( \frac{1}{6} \left ( 2\Delta \amp_{00}(s)+\Delta \amp_{20}(s)
  \right ) 
  \right . \right . \notag\\ &&&& &  \left . \left .
  -\frac{5}{3}\left (2 \Delta \amp_{02}(s)+\Delta \amp_{22}(s)
  \right )
  \right ) \frac{g_{\mu_1\mu_2}g_{\mu_3\mu_4}}{s^2} \right .
  \notag\\
  &&&& &+ \left ( 5\Delta \amp_{02}(s)-\frac{3}{2}\Delta
    \amp_{11}(s)+\frac{5}{2} \Delta \amp_{22}(s) \right )\frac{g_{\mu_1\mu_3}g_{\mu_2\mu_4}}{s^2}
  \notag\\
  &&&& & \left . + \left ( 5\Delta \amp_{02}(s)+\frac{3}{2}\Delta
    \amp_{11}(s)+\frac{5}{2} \Delta \amp_{22}(s) \right )
  \frac{g_{\mu_1\mu_4}g_{\mu_2\mu_3}}{s^2}
    \right ]
    \\
   Z_{\mu_1}Z_{\mu_2}\rightarrow Z_{\mu_3}Z_{\mu_4}
   &:& & \frac{g^4 v^4}{4c_w^4} & &\left [
  \left ( \frac{1}{3} \left ( \Delta \amp_{00}(s)+2\Delta \amp_{20}(s)
  \right )
	  \right . \right . \notag\\ &&&& &  \left . \left .
	  -\frac{10}{3}\left ( \Delta \amp_{02}(s)+2\Delta
    \amp_{22}(s) \right )
    \right ) \frac{g_{\mu_1\mu_2}g_{\mu_3\mu_4}}{s^2} \right .
  \notag\\
  &&&& & \left .+ 5 \left (\Delta \amp_{02}(s)+2\Delta \amp_{22}(s) \right )
  \frac{g_{\mu_1\mu_3}g_{\mu_2\mu_4}+ g_{\mu_1\mu_4}g_{\mu_2\mu_3}}{s^2} \right ]
\end{alignat}

\bibliographystyle{unsrt}

\end{document}